\UseRawInputEncoding
\documentclass[reprint, twocolumn, superscriptaddress]{revtex4-1}
\usepackage{bm, amsmath, amsfonts, amssymb, braket}
\usepackage{subfigure}
\usepackage{color}
\usepackage{graphicx}
\usepackage{dcolumn} 

\newcommand{\ii}{\text{i}}

\usepackage{rotating} 

\begin{document}
\title{Quantum anomaly, non-Hermitian skin effects, and entanglement entropy in open systems}
\author{Nobuyuki Okuma}
\email{okuma@hosi.phys.s.u-tokyo.ac.jp}
\author{Masatoshi Sato}
\affiliation{%
 Yukawa Institute for Theoretical Physics, Kyoto University, Kyoto 606-8502, Japan
}%

\date{\today}
\begin{abstract}
We investigate the roles of non-Hermitian topology in spectral properties and entanglement structures of open systems. In terms of spectral theory, we give a unified understanding of two interpretations of non-Hermitian topology: quantum anomaly and non-Hermitian skin effects, in which the bulk spectra extremely depend on the boundary conditions.
In this context, the fact that the intrinsic higher-dimensional skin effects under the full open boundary condition need the presence of the topological defects is understood in terms of the anomalous fermion production such as the Rubakov-Callan effect in the presence of the magnetic monopole. By using the unified interpretation, we classify the symmetry-protected and higher-dimensional skin effects.
In terms of the entanglement structure, we investigate steady states of fermionic open systems whose Liouvillian (rapidity) spectra host non-Hermitian topology.
We analyze dissipation-driven Majorana steady states in zero-dimensional open systems and relate them to the Majorana edge modes of topological superconductors by using the entanglement entropy. 
We also analyze a steady state of a one-dimensional open Fermi system with a non-Hermitian topological spectrum and relate it to the chiral edge states of the Chern insulator on the basis of the trace index defined from the entanglement spectrum. This correspondence indicates that the entanglement generates circular nonreciprocal currents under the periodic boundary condition and the skin-effect voltage with fermion accumulation under the open boundary condition. 
Finally, we discuss several related topics such as pseudospectral behaviors of Liouvillian dynamics and skin effects in interacting systems.

\end{abstract}
\maketitle
\section{Introduction}
Recently, non-Hermitian matrices have been extensively studied in various fields of physics.
In terms of the spectral theory, non-Hermiticity breaks lots of properties that hold in the Hermitian physics \cite{Bender-98, Bender-02, Bender-review, Konotop-review, Christodoulides-review, Gong-18, KSUS-19,ZL-19,ashida-gong-20,Poli-15,Zeuner-15,Zhen-15,Zhou-18,Weimann-17,Xiao-17,St-Jean-17,Bahari-17,Harari-18,Bandres-18,Zhao-19,Kozii-17,Yoshida-18,Yoshida-19,Bergholtz-19,Kimura-19,Okugawa-19,Budich-19,KBS-19,Bardarson-19,Ryu-20,Ye-20,Hatano-Nelson-96,Hatano-Nelson-97,Lee-16,MartinezAlvarez-18,Torres-2019,YW-18-SSH,YSW-18-Chern,Kunst-18,YM-19,KOS-20,YZFH-19,Lee-19,OKSS-20, Zhang-19,Okuma-19,Borgnia-19,Brandenbourger-19-skin-exp,Ghatak-19-skin-exp,Helbig-19-skin-exp, Hofmann-19-skin-exp,Xiao-19-skin-exp,Weidemann-20-skin-exp,Brandenbourger-19-skin-exp,Ghatak-19-skin-exp,Xiao-19-skin-exp,Weidemann-20-skin-exp,Trefethen,Okuma-Sato-20,Makris-08, Klaiman-08, Guo-09, Ruter-10, Lin-11, Regensburger-12, Peng-14, Wu-19, Yamamoto-19, Xiao-18,Song-yao-wang-19,Lieu-19,haga-20,Kozii-17,Yoshida-18,Yoshida-19,Michishita-20,Gong-18,Lee-14, Nakagawa-18, Li-19, Wu-19,Lindblad,Minganti-18,Lee-Vishwanath-19,Bessho-Sato-20,Prosen-2008}. 
For example, bulk eigenspectra of non-Hermitian matrices defined on lattice systems can strongly depend on the boundary conditions, while that of the Hermitian ones do not under the infinite-volume limit.
This phenomenon called the non-Hermitian skin effect \cite{Hatano-Nelson-96,Hatano-Nelson-97,Lee-16,MartinezAlvarez-18,Torres-2019,YW-18-SSH,YSW-18-Chern,Kunst-18,YM-19,KOS-20,YZFH-19,Lee-19,OKSS-20, Zhang-19,Okuma-19,Borgnia-19,Brandenbourger-19-skin-exp,Ghatak-19-skin-exp,Helbig-19-skin-exp, Hofmann-19-skin-exp,Xiao-19-skin-exp,Weidemann-20-skin-exp} has been extensively studied. The spectral theory of the non-Hermitian skin effect has been sophisticated in terms of the boundary-localized modes called non-Bloch wavefunctions \cite{YW-18-SSH,YSW-18-Chern,Kunst-18,YM-19,KOS-20,YZFH-19}, while the topological theories about its mathematical origin \cite{Lee-19,OKSS-20,Zhang-19} and symmetry-protected variants \cite{Okuma-19,OKSS-20} have also been developed.
Another example is the exceptional point \cite{Kozii-17,Yoshida-18,Yoshida-19,Bergholtz-19,Kimura-19,Okugawa-19,Budich-19,KBS-19}, where the non-Hermitian matrix is not diagonalizable, while Hermitian matrices can always be diagonalized.
In both examples, the origin of the exotic behaviors is the nonnormality of the matrix $H$ (i.e., $[H,H^\dagger]\neq0$) \cite{Trefethen,Okuma-Sato-20}.

Non-Hermitian matrices play important roles both in classical \cite{Makris-08, Klaiman-08, Guo-09, Ruter-10, Lin-11, Regensburger-12, Peng-14} and quantum systems \cite{Gong-18,Lee-14, Nakagawa-18, Li-19, Wu-19, Yamamoto-19, Xiao-18,Song-yao-wang-19,Lieu-19,haga-20,Kozii-17,Yoshida-18,Yoshida-19,Michishita-20}. 
For instance, the Lindblad formalism \cite{Lindblad} with the postselction \cite{Gong-18,ashida-gong-20} or the Green's function formalism \cite{Kozii-17,Yoshida-18,Yoshida-19,Michishita-20} enable one to define quantum systems described by the non-Hermitian Hamiltonians.
In addition, the Liouvillian quantum dynamics can be regarded as a liniear equation with a non-Hermitian matrix \cite{Minganti-18,Lieu-19}.
Although the spectral properties of quantum systems have been well investigated, the properties of the corresponding quantum states are still unclear.
In particular, it is known that the steady-state properties are not solely determined by the Liouvillian spectrum \cite{Lieu-19}.

In this paper, we investigate the steady-state properties of the open systems by using the topology of non-Hermitian spectra. We focus on the spectral properties and entanglement structures in the so-called AZ$^\dagger$ symmetry class \cite{KSUS-19}, which is essentially important for various quantum systems \cite{Lieu-19,Yoshida-19}. 
In terms of spectral theory, we give a unified understanding of two interpretations of the non-Hermitian topology: the quantum anomaly \cite{Lee-Vishwanath-19,Bessho-Sato-20} and the skin effects \cite{OKSS-20}.
In particular, we find the relationship between the higher-dimensional skin effects \cite{OKSS-20} and anomalous fermion production under the topological defects such as the Rubakov-Callan effect induced by magnetic monopoles \cite{Rubakov,callan}.
In terms of the steady-state properties, we investigate quadratic open Fermi systems \cite{Prosen-2008,Lieu-19}. 
Although the steady-state properties are not solely determined by the Liouvillian spectrum \cite{Lieu-19} as mentioned above, we identify a role of the non-Hermitian topology of the Liouvillian spectrum in the steady state under certain conditions.
In particular, we relate the zero-dimensional Majorana-fermion system and the one-dimensional complex-fermion system to the boundary states of the topological superconductor and the Chern insulator, respectively.
These results indicate that the non-Hermitian topology is important not only for the spectral properties but also for the quantum states in open quantum systems.

This paper is organized as follows.
In Sec. \ref{quadopenfermi},
we review the Lindblad equation and Liouvillian spectrum. In particular, we focus on the quadratic Fermi systems \cite{Prosen-2008}.
By introducing the $third$ $quantization$ \cite{Prosen-2008} for such systems, one can decompose the Liouvillian eigenvalues into the eigenvalues of a quadratic non-Hermitian matrix $Z$ called rapidity spectrum \cite{Lieu-19}. We also review the symmetry class of the non-Hermitian matrix $Z$ \cite{Lieu-19} called the AZ$^\dagger$ class \cite{KSUS-19}.
In Sec. \ref{physint},
we review two physical interpretations of the non-Hermitian topology in the AZ$^\dagger$ class and give a new interpretation that combines them.
We first define the ``non-Hermitian topology" used in this paper and review the AZ$^\dagger$ classification, which is identical to that of the anomalous gapless modes in the corresponding Hermitian class.
We then review the non-Hermitian skin effects and their topological origin. Relating the AZ$^\dagger$ classification to the skin effects, we give a unified understanding of the quantum anomaly and skin effects.
In Sec. \ref{zerodim}, we investigate the steady-state quantum entanglement of the zero-dimensional open Fermi systems, especially about the dissipation-driven Majorana fermion. As mentioned in Ref. \cite{Lieu-19}, the steady-state properties are not solely determined by the rapidity spectrum. We choose appropriate setups and relate them to the Majorana bound states of the one-dimensional topological superconductors.
In Sec. \ref{onedim}, we investigate the steady-state quantum entanglement of the one- and higher-dimensional open Fermi systems. We mainly consider the one-dimensional system whose rapidity spectrum is described by the Hatano-Nelson model.
We again choose the appropriate setups and relate it to the chiral edge state of the Chern insulator in terms of the entanglement spectrum.
In Sec. \ref{discussion}, we discuss several related topics.

\section{Quadratic open Fermi System and AZ$^{\dagger}$ Class \label{quadopenfermi}}
In this section, we review open quantum systems of free fermions described by the Lindblad equation.
First we introduce the Lindblad equation and corresponding Liouvillian superoperator.
Next, we define the quadratic open Fermi systems \cite{Prosen-2008}, the main targets of this paper.
By introducing the $third$ $quantization$ defined in Ref. \cite{Prosen-2008}, one can decompose the Liouvillian spectrum, which determines the dynamical properties of the Lindblad equation, into the ``one-particle" spectrum of a non-Hermitian matrix (rapidity spectrum \cite{Prosen-2008}).
Finally, we introduce the symmetry class of the quadratic Liouvillian \cite{Lieu-19}, which is identical to the AZ$^{\dagger}$ class introduced in Ref. \cite{KSUS-19}.

\subsection{Lindblad equation}
In this paper, we consider open quantum systems that consist of the system and the environment.
The starting point of this paper is the Lindblad equation \cite{Lindblad}:
\begin{align}
    \frac{d\rho}{dt}&=-i[H,\rho]+\sum_{\mu}\left(2L_\mu\rho L^\dagger_\mu-\{L^\dagger_\mu L_\mu,\rho \} \right),
\end{align}
where $H$ and $\rho$ are $s\times s$ matrices representing the  Hamiltonian and the density matrix of the system with $s$ being the Hilbert-space dimension. Couplings between the system and the environment are represented by $s\times s$ matrices $L_\mu$. 
The first and second terms represent the unitary and nonunitary dynamics of the density matrix, respectively.
Since the Lindblad equation is linear in $\rho$, it is convenient to introduce the Liouvillian superoperator $\mathcal{L}$:
\begin{align}
    \frac{d\rho}{dt}=\mathcal{L}\rho.\label{lioudyn}
\end{align}
By regarding the density matrix $\rho$ as a vector that consists of $s^2$ matrix elements $\rho_{i,j}$, $\mathcal{L}$ is represented as a $s^2\times s^2$ matrix whose elements are given by
\begin{align}
    \mathcal{L}_{ij,kl}:=&\sum_{\mu}2L_{\mu;i,k}L^\dagger_{\mu;l,j}
    -i(H-i\sum_{\mu}L_\mu^\dagger L_\mu)_{i,k}\delta_{l,j}\notag\\
    &+i(H+i\sum_{\mu}L_\mu^\dagger L_\mu)_{l,j}\delta_{ik}.
\end{align}
These representations enable one to treat the Lindblad equation as a linear equation. In other words, the dynamics of the system can be understood in terms of the eigenvalue problem of the Liouvillian matrix:
\begin{align}
    \mathcal{L}\rho^{(i)}=\lambda_i\rho^{(i)},\label{eigenproblem}
\end{align}
where $\lambda_i$ are known to satisfy Re$[\lambda_i]\leq0$ \cite{Minganti-18}.
A time-independent Liouvillian has at least one steady state \cite{Minganti-18}:
\begin{align}
    \mathcal{L}\rho_{\rm ss}=0.
\end{align}
If there is a unique steady state, the time-dependence of the density matrix is calculated as 
\begin{align}
    \rho(t)=\rho_{\rm ss}+\sum_{i\neq0}c_ie^{\lambda_i t}\rho^{(i)}.
\end{align}

\subsection{Liouvillian of quadratic open Fermi system}
Let us consider a system consists of $n$ complex fermions.
Since the size of the Hilbert space of the system is $2^n$, the Liouvillian is given by a $2^{2n}\times2^{2n}$ matrix, whose eigenvalue problem is numerically expensive for large $n$.
In the case of quadratic open Fermi systems defined below, however, the Liouvillian eigenspectra can be efficiently obtained \cite{Prosen-2008}.

A quadratic open Fermi system is a system whose Hamiltonian/bath operators consist of quadratic/linear terms of $n$ complex fermions ($2n$ Majorana fermions) \cite{Prosen-2008}: 
\begin{align}
    H=\sum^{2n}_{i,j=1}\gamma_iH_{i,j}\gamma_j,~L_\mu=\sum^{2n}_{i=1}l_{\mu,i}\gamma_i,
\end{align}
where $\gamma_{i}$ are Majorana fermions satisfying $\{\gamma_i,\gamma_j\}=2\delta_{ij}$. The Hamiltonian matrix is chosen to be an antisymmetric matrix $H^T=-H$.
Prosen has introduced the $third$ $quantization$ of the quadratic open Fermi system \cite{Prosen-2008}. In the $third$ $quantization$, the Liouvillian superoperator is expressed as a quadratic form of the $2n$ complex fermions ($4n$ Majorana fermions) \cite{Prosen-2008,Lieu-19}:
\begin{align}
    \mathcal{L}=\frac{2}{i}
    \begin{pmatrix}
    \bm{c}^{\dagger}& \bm{c}
    \end{pmatrix}
    \begin{pmatrix}
    -Z^T& Y\\
    0&Z
    \end{pmatrix}
    \begin{pmatrix}
    \bm{c}\\
    \bm{c}^{\dagger}
    \end{pmatrix},\label{quadliou}\\
    Z=H+i~{\rm Re} [M],~
    Y=2~{\rm Im}[M],\label{znonherm}
\end{align}
where $M_{i,j}:=\sum_{\mu}l_{\mu,i}l^*_{\mu,j}$ represents dissipation, and $\bm{c}=(c_1,\cdots,c_{2n})$ are $third$ $quantized$ complex fermions.
Note that the number of the complex fermions is doubled because of the degree of the physical bra and ket states of the density matrix.
In this formalism, the density matrix is a state lives on the doubled Hilbert space, while the Liouvillian superoperator behaves as an operator acts on the doubled Hilbert space.
\subsection{Rapidity and Liouvillian spectra}
By solving the eigenvalue problem of Eq. ($\ref{quadliou}$), we finally obtain \cite{Prosen-2008,Lieu-19}
\begin{align}
     \mathcal{L}=\frac{4}{i}\sum^{2n}_{i=1}E_i\bar{\beta}^{\dagger}_i\beta_i,
\end{align}
where $\{E_i \}$ is the eigenspectrum of $-Z$, or the rapidity spectrum in Ref. \cite{Prosen-2008} except for constant $-i/2$.
The anticommutation relations of fermionic operators are given by $\{\beta_i,\beta_j\}=\{\bar{\beta}^\dagger_i,\bar{\beta}^\dagger_j\}=0$ and $\{\bar{\beta}^\dagger_i,\beta_j\}=\delta_{ij}$.
Note that the creation operators need not to be the Hermitian conjugate of annihilation operators because the matrix in Eq. ($\ref{quadliou}$) is not always a normal matrix, whose Hermitian conjugate commutes with itself such as Hermitian matrices. 

\begin{figure}[]
\begin{center}
　　　\includegraphics[width=8cm,angle=0,clip]{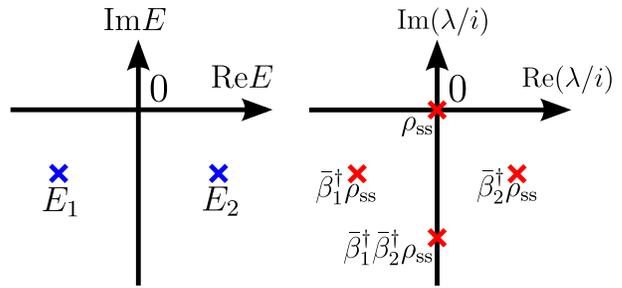}
　　　\caption{Rapidity (left) and Liouvillian (right) spectra of $n=1$ open Fermi system. In the $third$ $quantization$, $2^{2n}$ Liouvillian eigenstates are created by $2n$ complex fermions $\bar{\beta}^\dagger_i$'s.}
　　　\label{fig1}
\end{center}
\end{figure}

The Fock vacuum of the annihilation operators $\beta_i$ corresponds to the steady-state density matrix $\rho_{\rm ss}$ [i.e., $\beta_i\rho_{\rm ss}=0$].
The $2^{2n}$ eigenstates of the Liouvillian are constructed by creation operators $\{\bar{\beta}^\dagger_i\}$: 
\begin{align}
    \bar{\beta}_1^{\dagger\nu_1}\bar{\beta}^{\dagger\nu_1}_2\cdots\bar{\beta}^{\dagger\nu_{2n}}_{2n}\rho_{\rm ss},
\end{align}
where $\nu_i\in\{0,1\}$.
Thus, the Liouvillian eigenspectrum is given in terms of the rapidity spectrum (Fig.\ref{fig1}) by 
\begin{align}
\lambda=\frac{4}{i}\sum^{2n}_{i=1}E_i\nu_i.\label{rapliurel}
\end{align}
In summary, the dynamics of the quadratic open Fermi system is reduced to the ``one-particle" dynamics in the $third$ $quantization$. Note that the steady state is unique if and only if the rapidity spectrum does not contain $0$ \cite{Prosen-2008}.

\subsection{AZ$^\dagger$ symmetry class of rapidity spectrum}
As discussed above, the eigenvalue problem of the Liouvillian spectrum is related to that of a non-Hermitian matrix $Z$, which can be regarded as a generalization of free Hermitian Hamiltonians. In fact, it is identical to the Hamiltonian matrix in the absence of the dissipation. Here we discuss the generalization of Altland-Zirnbauer (AZ) symmetries \cite{Altland} of the non-Hermitian matrix $Z$ defined in Eq. (\ref{znonherm}). 

According to Ref. \cite{Lieu-19}, $Z$ is classified in the tenfold way that consists of the time-reversal symmetry (TRS), particle-hole symmetry (PHS), and chiral symmetry (CS):
\begin{align}
    {\rm TRS}&:~TZ^TT^{-1}=Z,~TT^*=\pm1,\notag\\
    {\rm PHS}&:~CZ^*C^{-1}=-Z,~CC^*=\pm1,\notag\\
    {\rm CS}&:~\Gamma Z^\dagger\Gamma^{-1}=-Z,~\Gamma^2=1,\label{AZdagger}
\end{align}
where $T,C,\Gamma$ are unitary matrices.
Note that the transpose is not equivalent to the complex conjugation in the presence of the non-Hermiticity, which indicates that other types of generalizations of the symmetry operations can be constructed by replacing $Z^T (Z^*)$ with $Z^* (Z^T)$, apart from the rapidity spectrum \cite{KSUS-19}. Nevertheless, the classification based on the symmetries ($\ref{AZdagger}$) is especially important in terms of not only the rapidity spectrum but also physical interpretations of the non-Hermitian topology discussed below.
In the following, we call Eq. ($\ref{AZdagger}$) as the AZ$^{\dagger}$ symmetries \cite{KSUS-19}, which is identical to the AZ symmetries in the absence of the non-Hermiticity, and focus on the topological phenomena in the AZ$^{\dagger}$ class.
For convenience, we include the classes A and AIII in the AZ$^{\dagger}$ class.

\section{Physical interpretations of Non-Hermitian topology in AZ$^{\dagger}$ class \label{physint}}
In this section, we review and reinterpret the non-Hermitian topological classification in the AZ$^{\dagger}$ class defined in Eq. (\ref{AZdagger}), which is relevant for various frameworks such as the quadratic open Fermi system, the main target of this paper, and the non-Hermitian effective Hamiltonian defined from the Green's function \cite{Kozii-17,Yoshida-18,Yoshida-19,Michishita-20}. We first review the non-Hermitian topological classification \cite{Gong-18,KSUS-19,ashida-gong-20} and its direct relation with the anomalous gapless modes in the Hermitian systems \cite{Lee-Vishwanath-19,Bessho-Sato-20}. We then review the topological origin of the non-Hermitian skin effect and its symmetry-protected and higher-dimensional variants \cite{OKSS-20}. We also remark an additional thought on the classification of skin effects, which is not mentioned in Ref. \cite{OKSS-20}. Finally, we give a unified understanding of the anomalous gapless modes and the skin effects.
In particular, we find the relationship between the higher-dimensional skin effects \cite{OKSS-20} and the anomalous fermion production under the topological defects including the Rubakov-Callan effect induced by the magnetic monopole \cite{Rubakov,callan}.
By using the unified interpretation, we classify the symmetry-protected and higher-dimensional skin effects.

\subsection{Non-Hermitian topology in AZ$^{\dagger}$ class}
\begin{table}[]
\caption{Non-Hermitian topological classification for AZ$^\dagger$ class (point-gap classification in Ref. \cite{KSUS-19}). 
The time-reversal, particle-hole, and chiral symmetries are defined in Eq. (\ref{AZdagger}).
This table is identical to the classification table of anomalous gapless modes in the corresponding Hermitian AZ class. For the class A, AI$^\dagger$, and AII$^\dagger$, the nontrivial groups also classify the non-Hermitian skin effects including higher-dimensional and symmetry-protected variants. }
\label{table1}
\centering
$$
\begin{array}{c|ccc|ccccccccc}
\mbox{~AZ}^\dagger~&T&C&\Gamma&0&1&2&3&4&5&6&7\\
\hline \hline
\textcolor{red}{{\rm A}}~(\mathcal{C}_1)&0&0&0&0&\textcolor{red}{\mathbb{Z}}&0&\textcolor{red}{\mathbb{Z}}&0&\textcolor{red}{\mathbb{Z}}&0&\textcolor{red}{\mathbb{Z}}\\ 
{\rm AIII}~(\mathcal{C}_0)&0&0&1&\mathbb{Z}&0&\mathbb{Z}&0&\mathbb{Z}&0&\mathbb{Z}&0\\
\hline
\textcolor{red}{{\rm AI}^\dagger}~(\mathcal{R}_7)&1&0&0&0&0&0&\textcolor{red}{2\mathbb{Z}}&0&\textcolor{red}{\mathbb{Z}_2}&\textcolor{red}{\mathbb{Z}_2}&\textcolor{red}{\mathbb{Z}}\\
{\rm BDI}^\dagger~(\mathcal{R}_0)&1&1&1&\mathbb{Z}&0&0&0&2\mathbb{Z}&0&\mathbb{Z}_2&\mathbb{Z}_2\\
{\rm D}^\dagger~(\mathcal{R}_1)&0&1&0&\mathbb{Z}_2&\mathbb{Z}&0&0&0&2\mathbb{Z}&0&\mathbb{Z}_2\\
{\rm DIII}^\dagger~(\mathcal{R}_2)&-1&1&1&\mathbb{Z}_2&\mathbb{Z}_2&\mathbb{Z}&0&0&0&2\mathbb{Z}&0\\
\textcolor{red}{{\rm AII}^\dagger}~(\mathcal{R}_3)&-1&0&0&0&\textcolor{red}{\mathbb{Z}_2}&\textcolor{red}{\mathbb{Z}_2}&\textcolor{red}{\mathbb{Z}}&0&0&0&\textcolor{red}{2\mathbb{Z}}\\
{\rm CII}^\dagger~(\mathcal{R}_4)&-1&-1&1&2\mathbb{Z}&0&\mathbb{Z}_2&\mathbb{Z}_2&\mathbb{Z}&0&0&0\\
{\rm C}^\dagger~(\mathcal{R}_5)&0&-1&0&0&2\mathbb{Z}&0&\mathbb{Z}_2&\mathbb{Z}_2&\mathbb{Z}&0&0\\
{\rm CI}^\dagger~(\mathcal{R}_6)&1&-1&1&0&0&2\mathbb{Z}&0&\mathbb{Z}_2&\mathbb{Z}_2&\mathbb{Z}&0\\
\hline
\hline
\end{array}
$$
\end{table}

References \cite{Gong-18,KSUS-19,ZL-19} have classified non-Hermitian matrices (``Hamiltonians") in two ways, i.e., the point-gap and line-gap classifications. In the latter scheme, the non-Hermitian Hamiltonians can be transformed to the Hermitian Hamiltonians without closing the gap and violating the symmetry constraints, which indicates that the classification table of them is identical to the Hermitian one \cite{Esaki-11,KSUS-19,ashida-gong-20}. 
In the following, we regard the former scheme as the non-Hermitian topological classification.

The (point-gap) non-Hermitian topological classification classifies a non-Hermitian Hamiltonian $H$ whose complex energy spectrum $\{E\}$ does not contain $E=0$ [i.e., $\det H\neq0$]. References \cite{Gong-18,KSUS-19,ZL-19} have shown the fact that the classification of $H$ is identical to that of the doubled Hermitian matrix $\tilde{H}$:
\begin{align}
\tilde{H} := 
\begin{pmatrix}
0&H\\
H^\dagger&0
\end{pmatrix}.
\end{align}
The time-reversal, particle-hole, and chiral symmetries of the doubled Hermitian Hamiltonian are expressed in terms of the AZ$^\dagger$ symmetries of the original non-Hermitian Hamiltonian $T$, $C$, and $\Gamma$, respectively:
\begin{align}
    {\rm TRS}&:~\tilde{T}\tilde{H}^*\tilde{T}^{-1}=\tilde{H},~\tilde{T}\tilde{T}^*=\pm1,~\tilde{T}:=
    \begin{pmatrix}
    0&T\\
    T&0
    \end{pmatrix},
    \notag\\
    {\rm PHS}&:~\tilde{C}\tilde{H}^*\tilde{C}^{-1}=-\tilde{H},~\tilde{C}\tilde{C}^*=\pm1,~\tilde{C}:=
    \begin{pmatrix}
    C&0\\
    0&C
    \end{pmatrix},\notag\\
    {\rm CS}_0&:~\tilde{\Gamma}_0\tilde{H}\tilde{\Gamma}_0^{-1}=-\tilde{H},~\tilde{\Gamma}_0^2=1,~\tilde{\Gamma}_0:=
    \begin{pmatrix}
    0&\Gamma\\
    \Gamma&0
    \end{pmatrix}.
\end{align}
In addition, there arises a chiral symmetry owing to the doubling process:
\begin{align}
    {\rm CS}&:~\tilde{\Gamma}\tilde{H}\tilde{\Gamma}^{-1}=-\tilde{H},~\tilde{\Gamma}^2=1,~\tilde{\Gamma}:=
    \begin{pmatrix}
    1&0\\
    0&-1
    \end{pmatrix}.\label{chiralsym}
\end{align}
Mathematically, the topological classification of the Hermitian Hamiltonians is performed by identifying the classifying space \cite{Schnyder-08,Schnyder-Ryu-review,Kitaev-09}.
The additional chiral symmetry that arises from the doubling process shifts the classifying space from $\mathcal{C}_i/\mathcal{R}_i$ to $\mathcal{C}_{i-1}/\mathcal{R}_{i-1}$, where $\mathcal{C}_{-1}=\mathcal{C}_{1}$ and $\mathcal{R}_{-1}=\mathcal{R}_{7}$.
Thus the non-Hermitian topological classification table in the AZ$^\dagger$ class is reduced to the Hermitian one in the corresponding class shifted by one line (Table \ref{table1}).

\subsection{Anomalous gapless mode and non-Hermitian topology}
References \cite{Lee-Vishwanath-19} has noticed the fact that the classification of the non-Hermitian topology (Table \ref{table1}) is identical to that of anomalous gapless modes or quantum anomalies of free fermion systems, which cannot exist in isolated Hermitian lattice systems but can exist in the $d$-dimensional boundaries of the $(d+1)$-dimensional Hermitian lattice systems. 
In the same context, Ref. \cite{Bessho-Sato-20} has shown the correspondence between the point-gap classification and the (real) line-gap classification, which classifies the anomalous gapless modes in the AZ$^\dagger$ class or those in the corresponding Hermitian AZ class, and summarized it as the generalized Nielsen-Ninomiya theorem. 
We here review this correspondence in the class A in one dimension.

\begin{figure*}[]
\begin{center}
　　　\includegraphics[width=17cm,angle=0,clip]{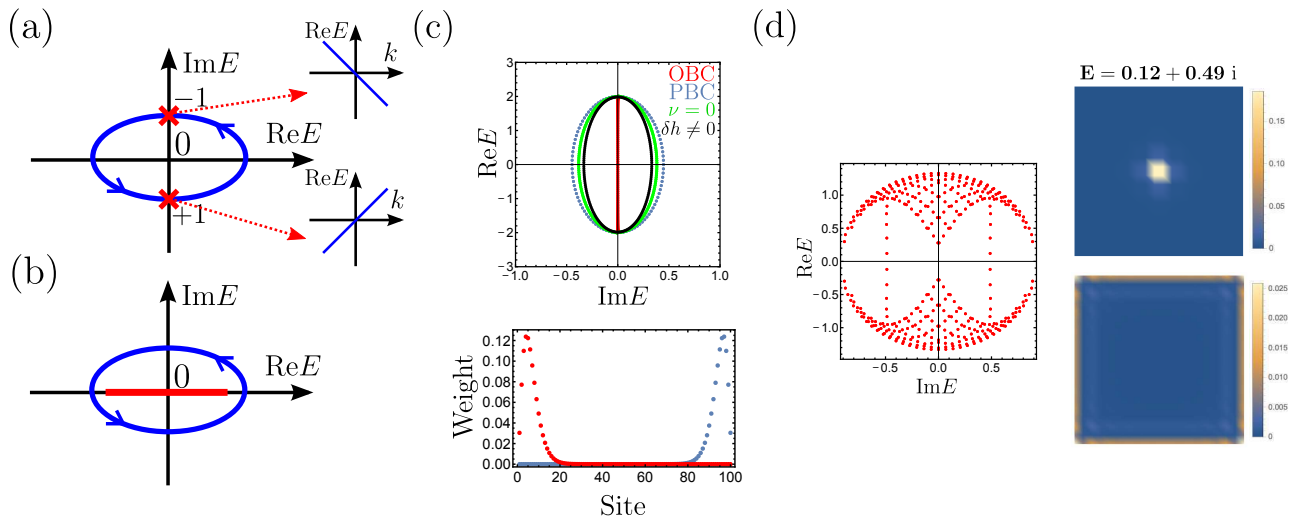}
　　　\caption{Non-Hermitian topology and its physical interpretations. (a) Correspondence between the non-Hermitian topology and the number of anomalous gapless modes (Re$E=0$).
　　　The PBC spectral curve with the winding number counts the topological charge of anomalous gapless modes. (b) Topological aspect of the non-Hermitian skin effect. The PBC spectral curve (blue) with nonzero winding indicates that the OBC spectrum (red) is drastically different from the PBC one. (c) Spectra and wavefunctions at $E=1.948$ of one-dimensional $\mathbb{Z}_2$ skin effect protected by the AII$^\dagger$ time-reversal symmetry, adopted from \cite{OKSS-20}. For $\nu=1$, the OBC spectrum (red) is drastically different from the PBC one (blue), while for $\nu=0$ or $\delta h\neq0$, the skin effect is suppressed. Each $\mathbb{Z}_2$ skin mode has the Kramers counterpart localized at the other edge. (c) Spectra and wavefunctions of two-dimensional $\mathbb{Z}_2$ skin effect protected by the AII$^\dagger$ time-reversal symmetry, adopted from the supplemental material of \cite{OKSS-20}. Only $\mathcal{O}(L)$ modes in $\mathcal{O}(L^2)$ modes are localized at the boundary and the topological defect ($\pi$ flux). }
　　　\label{fig2}
\end{center}
\end{figure*}

In general, the one-dimensional class-A non-Hermitian topological number is given by the winding number $W(E=0)$ of the complex eigenspectral curve(s) around the origin $E=0$ \cite{Gong-18}, where $W(E\in\mathbb{C})\in\mathbb{Z}$ is defined as 
\begin{align}
    W(E):=\sum_a\int_{0}^{2\pi}\frac{dk}{2\pi i}\frac{d}{dk}\log(E_{k,a}-E).\label{wnumber}
\end{align}
Here we assume the periodic boundary condition (PBC), and $E_{k,a}$ is the complex energy dispersion 
with band indices $a$.
In the same context,
Ref. \cite{Bessho-Sato-20} has shown that this winding number counts the gapless zero modes (Re$E=0$) with Im$E>0$/Im$E<0$ :
\begin{align}
    &W(0)=-\sum_{a,\alpha:~\mathrm{Im}E>0}\nu_{a,\alpha}=\sum_{a,\alpha:~\mathrm{Im}E<0}\nu_{a,\alpha},\label{nielsen}\\
    &\nu_{a,\alpha}=\mathrm{sgn}~(\mathrm{Re}~[dE_k/dk]_{k=k_{a,\alpha}}),
\end{align}
where the chirality $\nu_{a,\alpha}=\pm 1$ is assigned for the right-/left-moving chiral zero mode at $k=k_{a,\alpha}$. The simplest example is the Hatano-Nelson model \cite{Hatano-Nelson-96,Hatano-Nelson-97} without disorders under the PBC:
\begin{align}
    H_{\rm HN}=\sum_{i}\left[(t+g)c^\dagger_{i+1}c_i+(t-g)c^\dagger_i c_{i+1}\right],\label{hatano}
\end{align}
where $(c,c^\dagger)$ are the bosonic/fermionic annihilation and creation operators, $t\in\mathbb{R}$ is the Hermitian hopping, and $g\in\mathbb{R}$ is the non-Hermitian asymmetric hopping.
The complex energy spectrum is given by $\{E_k=(t+g)e^{ik}+(t-g)e^{-ik}~|~k\in[0,2\pi)\}$, which forms an ellipse in the complex plane.
Thus, the non-Hermitian topological number of $H_{\rm HN}$ is given by $W(0)=\mathrm{sgn}(t)~\mathrm{sgn}(g)$.
The formula (\ref{nielsen}) holds in this model [Fig.\ref{fig2} (a)], for example for $t>0$ and $g>0$, $W(0)=1$, $\nu_{k=\pi/2}=-1$ with $E= 2ig$, and $\nu_{k=-\pi/2}=1$ with $E= -2ig$.

Bessho and one of the present authors have shown that similar formulae for non-Hermitian topological numbers and topological charges of gapless modes in the general AZ$^\dagger$ class hold and summarized them as the generalized Nielsen-Ninomiya theorem \cite{Bessho-Sato-20}.
They have noticed the fact that any topological phases can be generated by a set of primitive models in the K-theory classification, which means that any topological properties including Eq. (\ref{nielsen}) and its generalization can be checked by investigating the primitive models.
They have used the primitive models given in Ref. \cite{Lee-Vishwanath-19} and proved the generalized Nielsen-Ninomiya theorem for those models.
In classical non-Hermitian systems, a nontrivial non-Hermitian topology of the ``Hamiltonian" indicates the presence of anomalous gapless modes that survive in the long-time dynamics.
In quantum systems, on the other hand, the physical consequence should depend on the situation.
In the later sections, we analyze the open Fermi systems with nontrivial rapidity spectra under certain conditions.

\subsection{Skin effects as non-Hermitian topological phenomena\label{skinastop}}
The relationship between the skin effect and the winding number has been noticed by Ref. \cite{Gong-18,Lee-19}.
The authors and collaborators have shown that the physics of the non-Hermitian skin effect is related to the physics of ``exact" zero modes of a Hermitian topological insulator \cite{OKSS-20,Okuma-Sato-20}.
By generalizing this idea to other classes, they have proposed the symmetry-protected and higher-dimensional variants \cite{OKSS-20}.
We here briefly review Ref. \cite{OKSS-20} and remark an additional thought on the AZ$^\dagger$ symmetry.
See Ref. \cite{OKSS-20} and Appendix \ref{appendixa} for details.

The non-Hermitian skin effect is a phenomenon in which the bulk eigenspectrum of a one-dimensional non-Hermitian lattice Hamiltonian under the open boundary condition (OBC) is drastically different from that under the PBC. 
In general, the OBC spectrum is given by replacing the Bloch wave factor $e^{ik}$ of the PBC dispersion $E(e^{ik})$ with the non-Bloch wave factor $\beta\in\mathbb{C}$, whose trajectory called the generalized Brillouin zone is determined via the non-Bloch band theory \cite{YW-18-SSH,YSW-18-Chern,Kunst-18,YM-19,KOS-20,YZFH-19}.
The skin effect is characterized by the localized "skin" wavefunctions with $|\beta|\neq1$.
For example, the OBC spectrum of the Hatano-Nelson model (\ref{hatano}) is given by replacing $e^{ik}$ of the PBC spectrum with $\beta=\sqrt{|t-g|/|t+g|}e^{ik}$ [Fig.\ref{fig2} (b)]. 

The key observation to understand this phenomenon is that the OBC spectral curve of a non-Hermitian Hamiltonian has no winding if there is no symmetry except for the U(1) charge conservation (class A in the AZ$^\dagger$ classification), while the PBC spectrum can have nontrivial winding as in the case of the Hatano-Nelson model [Fig.\ref{fig2} (b)]. 
Reference \cite{OKSS-20} has shown that in the class A, the non-Hermitian skin effect inevitably occurs if the PBC curve has a nonzero winding number (\ref{wnumber}), i.e., the class-A nontrivial non-Hermitian topological number in $\mathbb{Z}$, and the OBC spectrum cannot have a nonzero winding number.
This statement is best expressed in terms of the direct correspondence between the class-A non-Hermitian skin modes and the boundary ``exact" zero modes of the class-AIII Hermitian topological insulator, both of which are classified by the same winding number ($\mathbb{Z}$).
As in the topological classification, let us consider the doubled Hermitian Hamiltonian of $H-E$, where $E$ is an complex eigenenergy of $H$:
\begin{align}
\tilde{H}_E := 
\begin{pmatrix}
0&H-E\\
H^\dagger-E^*&0
\end{pmatrix}.\label{he}
\end{align}
If $H-E$ has nontrivial non-Hermitian topology ($W(E)\neq0$), there exist boundary exact zero modes constructed from the skin mode $|E\rangle$ and its left-eigenstate counterpart (i.e., $\langle\!\langle E|H=E\langle\!\langle E|$):
\begin{align}
    \tilde{H}_E
    \begin{pmatrix}
    0\\
    |E\rangle
\end{pmatrix}=0,~
    \tilde{H}_E
    \begin{pmatrix}
    |E \rangle\!\rangle\\
    0
\end{pmatrix}=0.
\end{align}
These two zero modes are localized at the different edges, which are nothing but the boundary zero modes of class-AIII topological insulators.
For example, the doubled Hermitian Hamiltonian of the Hatano-Nelson model is the Su-Schrieffer-Heeger model \cite{SSH-79}, a typical one-dimensional class-AIII Hermitian topological insulator with boundary zero modes.
This symmetry shift from the class A to the class AIII stems from the chiral symmetry (\ref{chiralsym}).
Note that this correspondence does not hold for quasi-zero modes of topological insulators, which become exact zero modes only in the thermodynamic limit. Instead, such quasi-zero eigenvalues of topological insulators correspond to the pseudospectra of the non-Hermitian Hamiltonian in a similar manner (see Ref. \cite{Okuma-Sato-20}).

The above correspondence implies that the non-Hermitian skin effect can be generalized to other symmetry classes and dimensions by considering the correspondence between non-Hermitian skin modes and topological exact zero modes.
In particular, Ref. \cite{OKSS-20} has investigated the one- and two-dimensional $\mathbb{Z}_2$ skin effects protected by the time-reversal symmetry, which correspond to the AII$^{\dagger}$ classification (Table \ref{table1}), while Ref.\cite{Terrier-20} has investigated the three-dimensional one in a different context.
The doubled Hermitian Hamiltonian for the one-dimensional class-AII$^\dagger$ skin effect is a class-DIII topological superconductor, which has Majorana doublets as the topological zero modes at both edges:
\begin{align}
    \begin{pmatrix}
    0\\
    |E\rangle
\end{pmatrix},~
    \begin{pmatrix}
    T|E\rangle^*\\
    0
\end{pmatrix},~
    \begin{pmatrix}
    0\\
    T|E\rangle\!\rangle^*
\end{pmatrix},~
 \begin{pmatrix}
    |E\rangle\!\rangle\\
    0
\end{pmatrix}.
\end{align}
Correspondingly, the non-Hermitian skin mode $|E\rangle$ has the Kramers counterpart $T|E\rangle\!\rangle^*$ localized at the other edge.
As an example, let us consider the stack of the Hatano-Nelson model with the opposite asymmetric hopping under the mixing term $\Delta$ that preserves time-reversal symmetry:
\begin{align}
H \left( k \right)
= \left( \begin{array}{@{\,}cc@{\,}} 
	H_{\rm HN} \left( k \right) & 2\Delta \sin k \\
	2\Delta \sin k & H_{\rm HN}^{T} \left( -k \right) \\ 
	\end{array} \right),\label{z2skin}
\end{align}
where we give the model in the momentum space.
The corresponding spectra and eigenmodes are shown in Fig.\ref{fig2}(c) adopted from Ref. \cite{OKSS-20}. If the $\mathbb{Z}_2$ index \cite{KSUS-19,OKSS-20} $\nu=1$, the $\mathbb{Z}_2$ skin effect occurs. For $\nu=0$ constructed from two copies of the model ($\ref{z2skin}$) with coupling terms between them, on the other hand, the $\mathbb{Z}_2$ skin effect is suppressed owing to the coupling term between Kramers pairs localized at the different edges.
Also, even a small perturbation $\delta h$ that does not preserve the time-reversal symmetry destroys the $\mathbb{Z}_2$ skin effect. In the infinite-volume limit, such a perturbation can be taken as an infinitesimally small value.
The authors have investigated this infinitesimal instability for the model obtained by setting $\Delta=0$ \cite{Okuma-19}.
Remarkably, the order of limits changes the physics as in the case of the spontaneous symmetry breaking \cite{Altland-Simons}:
\begin{align}
\lim_{\delta h\rightarrow0}\lim_{n\rightarrow\infty}\neq\lim_{n\rightarrow\infty}\lim_{\delta h\rightarrow0}.
\end{align}
The left-hand side corresponds to the infinitesimal instability.
The $\mathbb{Z}_2$ nature and the infinitesimal instability against the symmetry-breaking perturbation indicate the symmetry-protected topological nature of the $\mathbb{Z}_2$ skin effect.

More nontrivial phenomena occur in two dimensions. Reference \cite{OKSS-20} has investigated the following two-dimensional model of the two-dimensional $\mathbb{Z}_2$ skin effect protected by the time-reversal symmetry in the supplemental material:
\begin{align}
H \left( \bm{k} \right)=\sin k_x \sigma_x+\sin k_y\sigma_y+i\Gamma(\cos k_x+\cos k_y)\sigma_0,\label{2dskin}
\end{align}
where $\sigma_\mu$ are the Pauli matrices ($\mu=0$: identity matrix).
This model consists of the Hermitian Dirac Hamiltonian and the valley-dependent dissipation $\Gamma\in\mathbb{R}$.
The corresponding doubled Hermitian Hamiltonian describes a class-DIII topological superconductor.
Among several boundary conditions investigated in Ref. \cite{OKSS-20}, the open boundary condition in both $x$ and $y$ directions (full OBC) is essentially important. Under this boundary condition, the skin effect does not occur, which can be explained in terms of the well-known fact that the two-dimensional class-DIII topological superconductor, which corresponds to the doubled Hermitian Hamiltonian in the present case, cannot have the ``exact" Majorana zero modes under the full OBC, and there should be $\mathcal{O}(1/L)$ finite gap with $L$ being the system size.
As noted above, the correspondence between skin-modes and zero modes does not hold for such quasi-zero modes, which leads to the absence of the skin effect.
It is also well known that the exact zero modes can exist in the presence of the topological defect ($\pi$ flux) in a two-dimensional class-DIII topological superconductor \cite{Qi-09}. Thus, the $\mathbb{Z}_2$ skin effect occurs in the presence of the $\pi$ flux even under the full OBC [Fig.\ref{fig2} (d)]. Remarkably, only $\mathcal{O}(L)$ modes among the $\mathcal{O}(L^2)$ eigenstates are localized at the boundary or the $\pi$ flux, while all the eigenstates are localized at the boundary in the Hatano-Nelson model. In general, the ($d>1$)-dimensional skin effects under the full OBC need the topological defects, and not all the eigenstates but $\mathcal{O}(L)$ modes are localized, both of which come from the fact that only the exact zero modes of the topological boundary states correspond to the skin modes. Note that we here use the terminology ``higher-dimensional skin effects" as the skin effects protected by the non-Hermitian topology in the same dimension, and distinguish them from the skin effects in higher dimensions protected by lower-dimensional non-Hermitian topology. For example, the conventional skin effect can occur in the higher dimensions if the one-dimensional class-A non-Hermitian topology in one direction can be defined for $H(\bm{k}^{\parallel})$, where $\bm{k}^{\parallel}$ is the co-dimensional momentum.

Finally, we remark an additional thought on the classification of the skin effects, which was not mentioned in Ref. \cite{OKSS-20}.
In the construction of the doubled Hermitian Hamiltonian (\ref{he}), the additional constant $E$ can destroy the symmetry of the non-Hermitian Hamiltonian $H$, for example for the particle-hole symmetry in the AZ$^\dagger$ symmetry.
Among the various types of the fundamental symmetry operations raised in the 38-fold classification \cite{OKSS-20}, only the time-reversal symmetries in the AZ$^\dagger$ class are not broken by adding a constant because the AZ$^{\dagger}$ time-reversal pair ($|E\rangle$, $T|E\rangle\!\rangle^*$) shares the same complex eigenenergy.
Thus, the above discussion can be applied for the class A, AI$^\dagger$, and AII$^{\dagger}$ in any dimensions, which implies that these three classes are essential for the skin-effect classification (Table \ref{table1}).
Note that additional fundamental symmetries can give a constraint on the topological invariants of these three classes and the shape of the spectrum, which induces the new trivial or nontrivial classification group, while they might not correspond to the essentially different phenomena. 
Also, in the presence of an on-site unitary symmetry that commutes with the Hamiltonian, the skin effect can separately occur in each eigenspace of the symmetric operation. In Ref. \cite{Okuma-19}, the stack of the Hatano-Nelson model with spin-dependent asymmetric hopping terms that commutes with the $z$-component spin operator has been investigated, in which the skin effect is destroyed by an infinitesimally small perturbation that mixes the different spin sectors.

\subsection{Unified understanding of skin effects and anomalous gapless modes}

As discussed above, the non-Hermitian topology in the AZ$^\dagger$ class detects the anomalous gapless modes, while it also explains the origin of non-Hermitian skin effects in the class A, AI$^\dagger$, and AII$^\dagger$.
We here give another understanding of skin effects in terms of anomalous gapless modes, which unifies the arguments in Refs \cite{Lee-Vishwanath-19,Bessho-Sato-20,OKSS-20}.

Let us again consider the one-dimensional class-A non-Hermitian topology. For the nontrivial non-Hermitian topology, chiral modes with nonzero group velocity should survive in the long-time dynamics, at least in the classical systems mentioned above. Thus there exists the non-equilibrium current of chiral modes under the PBC. If we impose the OBC, on the other hand, such a transport phenomenon induces the accumulation of the modes at the boundaries, which leads to the class-A skin effect.

\begin{figure}[]
\begin{center}
　　　\includegraphics[width=8cm,angle=0,clip]{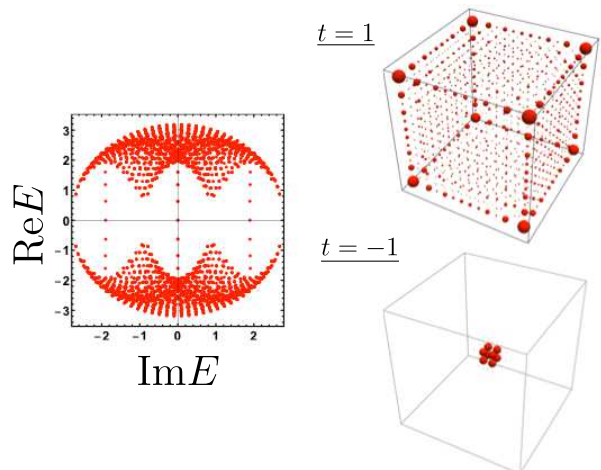}
　　　\caption{Spectrum and skin wavefunctions at $E=-0.00041i$ of three-dimensional class-A skin effect under the monopole gauge configuration on lattice with $n=14$. The model is given by Eq. (\ref{3dskin}). The model parameters are $t=\pm1$ and $\Gamma=0.5$. The skin wavefunction is localized at the surface/monopole for $t=\pm1$, while the spectrum does not depend on the sgn $t$.}
　　　\label{fig3}
\end{center}
\end{figure}

In general, the skin effects can be interpreted as a consequence of the anomaly-induced nonequilibrium current and its accumulation at the boundary or a defect.
In the one-dimensional $\mathbb{Z}_2$ skin effect protected by the AII$^\dagger$ symmetry, the accumulation occurs at both of the ends, which is similar to the spin accumulation in the spin Hall effect.
In the same context, the absence of the higher-dimensional skin effects under the full OBC with no topological defect is explained as the absence of the nonzero net current at the boundary.
In the presence of the topological defects, on the other hand, higher-dimensional skin effects under the full OBC occur, as we mentioned.
In terms of the anomalous gapless modes, it has been known that topological defects cause the fermion production \cite{Witten,Rubakov,callan,yamamoto-monopole}. Analogous phenomena occur in the higher-dimensional skin effects under the full OBC.
For example, the class-A three-dimensional skin effect under the topological defect corresponds to the Rubakov-Callan effect in the quantum field theory, which is expected to explain the monopole catalysis of the proton decay \cite{Rubakov,callan}. 
An explicit model of the class-A (also AII$^\dagger$) three-dimensional skin effect is given by
\begin{align}
    H(k)=t\sum^{3}_{i=1}\sin k_i \sigma_i+i\Gamma \sum^{3}_{i=1}\cos k_i,\label{3dskin}
\end{align}
where $t\in\mathbb{R}$ represents the kinetic energy.
We perform a numerical diagonalization under the full OBC and the magnetic monopole at the center of the system with $n=14$.
The configuration with the topological defect has not been investigated in Ref. \cite{Terrier-20}.
To describe the (unit) monopole magnetic field $\bm{B}=\hat{r}/2r^2$, we use the following gauge configuration and the Peierls phase from the site $i$ to the site $j$ \cite{monopolegauge}:
\begin{align}
    \bm{A}&=-\frac{1+\cos\theta}{2}\nabla\phi,\notag\\
    e^{i\theta_{i,j}}&=\exp\left[i\int^{j}_{i}d\bm{l}\cdot\bm{A}\right],
\end{align}
where $(\theta,\phi)$ is the spherical angle.
The spectrum and the skin wavefunctions at $E=-0.00041i$ are plotted in Fig.\ref{fig3}.
For $t=\pm1$, the wavefunction is localized at the surface/monopole.
This sign dependence indicates the $\mathbb{Z}$ nature of the class-A three-dimensional skin effect (Table \ref{table1}).
As in the case of the two-dimensional skin effect discussed above, the skin effect does not occur in the absence of the magnetic monopole.
The localization at the surface or the monopole is a non-Hermitian lattice analogue of the fermion production via the monopole catalysis.

In summary, one can construct non-Hermitian skin effects corresponding to the anomaly-related transport phenomena under considered symmetries, dimensions, and setups. In other words, the classification of the skin effects is nothing but that of the anomaly-related transport phenomena.

\begin{table}[]
\caption{Classification of non-Hermitian skin effects.
This table is identical to the classification table of intrinsic point gap topology in the AZ$^\dagger$ class \cite{OKSS-20,Shiozaki-prep}. A, AI$^\dagger$, and AII$^\dagger$ are the essential classes of non-Hermitian skin effects, and nontrivial groups in the other classes can be related to those in the essential three classes. Owing to the equivalence between the PHS in the AZ$^\dagger$ class and the TRS in the AZ class \cite{KSUS-19}, classification groups for the AI and AII classes are equivalent to those for the D$^\dagger$ and C$^\dagger$ classes, respectively. }
\label{table2}
\centering
$$
\begin{array}{c|ccccccccc}
\mbox{Symmetry}&0&1&2&3&4&5&6&7\\
\hline \hline
\textcolor{red}{{\rm A}}&0&\textcolor{red}{\mathbb{Z}}&0&\textcolor{red}{\mathbb{Z}}&0&\textcolor{red}{\mathbb{Z}}&0&\textcolor{red}{\mathbb{Z}}\\ 
{\rm AIII}&0&0&0&0&0&0&0&0\\
\hline
\textcolor{red}{{\rm AI}^\dagger}&0&0&0&\textcolor{red}{2\mathbb{Z}}&0&\textcolor{red}{\mathbb{Z}_2}&\textcolor{red}{\mathbb{Z}_2}&\textcolor{red}{\mathbb{Z}}\\
{\rm BDI}^\dagger&0&0&0&0&0&0&0&0\\
{\rm D}^\dagger/{\rm AI}&0&\mathbb{Z}&0&0&0&2\mathbb{Z}&0&0\\
{\rm DIII}^\dagger&0&\mathbb{Z}_2&\mathbb{Z}_2&0&0&0&0&0\\
\textcolor{red}{{\rm AII}^\dagger}&0&\textcolor{red}{\mathbb{Z}_2}&\textcolor{red}{\mathbb{Z}_2}&\textcolor{red}{\mathbb{Z}}&0&0&0&\textcolor{red}{2\mathbb{Z}}\\
{\rm CII}^\dagger&0&0&0&0&0&0&0&0\\
{\rm C}^\dagger/{\rm AII}&0&2\mathbb{Z}&0&0&0&\mathbb{Z}&0&0\\
{\rm CI}^\dagger&0&0&0&0&0&\mathbb{Z}_2&\mathbb{Z}_2&0\\
\hline
\hline
\end{array}
$$
\end{table}

\subsection{Classification of non-Hermitian skin effects}
We here discuss the classification of non-Hermitian skin effects under the fundamental symmetries.
As noted above, skin effects can be classified by investigating the anomaly-related transport.
This does not mean that all of the nontrivial groups in Table \ref{table1}, which describe anomalous gapless modes, classify the skin effects.
Apparently, the zero-dimensional classification does not describe the skin effects because of the absence of the boundaries.
In such a classification, all or some nontrivial matrices can be adiabatically connected to the anti-Hermitian topological insulators with keeping det $H\neq0$. Since anti-Hermitian matrices do not show the skin effects, one should get rid of such cases from the anomaly classification in order to obtain the skin-effect classification.
References \cite{OKSS-20,Shiozaki-prep} have introduced the intrinsic non-Hermitian point-gap topology for the 38 symmetry classes, which is defined by the point-gap groups divided by the subgroups that classify the Hermitian and anti-Hermitian topological insulators.
In the case of the AZ$^\dagger$ ten-fold classes, the nontrivial point-gap topology has no subgroup that can be connected to the Hermitian topological-insulator groups since it characterizes the Hermitian gapless modes.
Thus, the skin-effect classification is included in the intrinsic point-gap topological classification in the AZ$^\dagger$ class (Table \ref{table2}).

The above discussion narrows down the candidates of the skin-effect classification groups to the AZ$^\dagger$ intrinsic point-gap ones.
As noted in Sec. \ref{skinastop}, A, AI$^\dagger$, and AII$^\dagger$ are the essential classes of non-Hermitian skin effects because one can construct the direct correspondence between the skin modes and the boundary zero modes of topological insulators. Correspondingly, the point-gap classification is identical to the intrinsic point-gap classification in these three classes.
By comparing the generators of nontrivial groups in these three classes and those of the other intrinsic point-gap groups, we find that the skin-effect classification is given by the the AZ$^\dagger$ intrinsic point-gap classification itself.

\section{Non-Hermitian topology and entanglement in zero-dimensional open systems \label{zerodim}}
In this section, we analyze zero-dimensional open Fermi systems whose rapidity spectra have nontrivial non-Hermitian topology, especially the dissipative system of Majorana fermion, which is the anomalous zero mode of the particle-hole symmetry or quantum anomaly of the O($N$) rotational symmetry.
It is known that the steady-state properties are not solely determined by the rapidity spectral behaviors \cite{Lieu-19}.
Nevertheless, nontrivial rapidity spectra play an important role in understanding the quantum entanglement of the steady-state Majorana system. By choosing certain setups, we relate the steady-state Majorana fermions to the Majorana boundary zero modes of the topological superconductors.

\subsection{Background: Andreev vs. Majorana controversy}
The Majorana fermion, which is a fermion that is its own antiparticle, is one of the promising candidates of qubits for the topological quantum computation \cite{Nayak}. 
Such a fermion is expected to appear as the boundary zero modes of the topological superconductors and has been extensively studied in terms of topological physics \cite{Kitaev-01,Nayak}.
However, several papers have pointed out that nearly zero energy Andreev bound states in topologically trivial phase seem to reproduce lots of Majorana-like behaviors \cite{Kells-12,Parada-12,Liu-17,Moore-18,Moore-18-2,vuik-18,prada-20}.
References \cite{sanjose-16,avila-19} have proposed that the following non-Hermitian Hamiltonian defined from the Green's function explains such physics:
\begin{align}
    H=
    \begin{pmatrix}
    -i\Gamma^R_0&-i E_0\\
    i E_0&-i\Gamma^L_0
    \end{pmatrix},\label{majorananonham}
\end{align}
where $E_0$ is the fermion energy, and $\Gamma^{R/L}_0$ are the Majorana-dependent dissipation. Owing to the different lifetime of two Majorana fermions, the state is thought to be similar to the topological boundary Majorana fermion, though the properties such as the entanglement structure of the state is not so clear only from the spectral behavior.
In this section, we investigate this physics in terms of the rapidity spectrum and entanglement structure in the Lindblad physics.

\subsection{Rapidity spectrum and Majorana zero modes}
\begin{table}[t]
\caption{Zero-dimensional non-Hermitian topology and its physical interpretation.
The AZ$^\dagger$ classification is equivalent to the AZ Hermitian classification of the zero-energy bound states \cite{teo-kane}.}
\label{table3}
\centering
$$
\begin{array}{ccl}
\mbox{Symmetry}&\mbox{Classification}&\mbox{Interpretation}\\
\hline \hline
{\rm AIII}&\mathbb{Z}&\mbox{~Chiral Dirac~}\\
{\rm BDI}^\dagger&\mathbb{Z}&\mbox{~Chiral Majorana~}\\
{\rm D}^\dagger&\mathbb{Z}_2&\mbox{~Majorana~}\\
{\rm DIII}^\dagger&\mathbb{Z}_2&\mbox{~Majorana Kramers~}\\
{\rm CII}^\dagger&2\mathbb{Z}&\mbox{~Chiral Majorana Kramers~}\\
\hline
\hline
\end{array}
$$
\end{table}

\begin{figure}[]
\begin{center}
　　　\includegraphics[width=8cm,angle=0,clip]{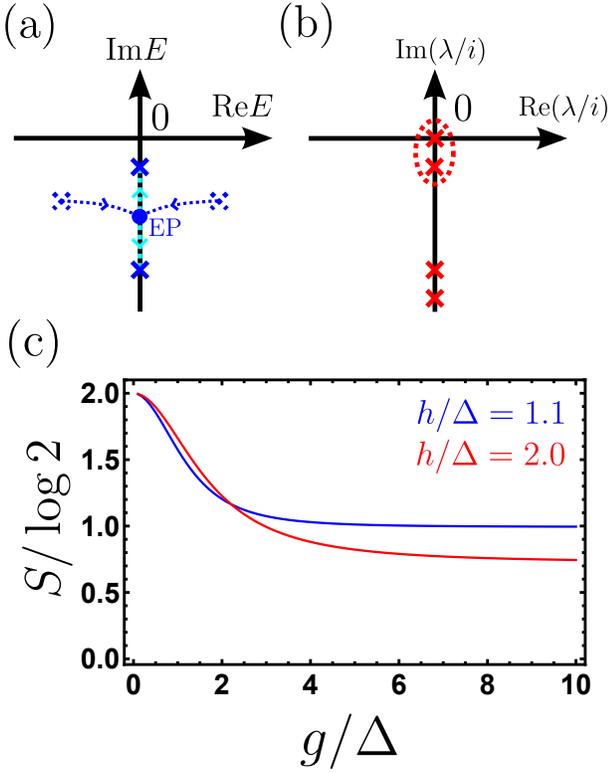}
　　　\caption{(a) Rapidity and (b) Liouvillian spectra of steady-state Majorana fermion system. In the rapidity spectrum, the topological phase transition occurs at the exceptional point. In the Liouvillian spectrum, the strong dissipation induces the steady-state quasi degeneracy. (c) Dissipation-dependence of entanglement entropy in zero-dimensional Majorana open Fermi system.}
　　　\label{fig4}
\end{center}
\end{figure}

In the following, we consider the zero-dimensional system of the Bogoliubov fermion $(\alpha,\alpha^\dagger)$, a superposition of the electron and hole sectors. Suppose that the loss and gain are described by the annihilation and creation of a fermion that consists only of the electron sector.
Such a system is described by the following Lindblad equation:
\begin{align}
    \frac{d\rho}{dt}=&-i[H,\rho]+(1-f(\epsilon))\left[2c\rho c^\dagger-\{c^\dagger c,\rho\}\right]\notag\\
    &+f(\epsilon)\left[2c^\dagger\rho c-\{cc^\dagger,\rho\}\right],\\
    c=&a\gamma_1+b\gamma_2,~\gamma_1=\alpha+\alpha^\dagger,~\gamma_2=i(\alpha-\alpha^\dagger),\\
    H=&-i\frac{\epsilon}{2} \gamma_1\gamma_2=\epsilon \alpha^\dagger\alpha+\mathrm{const.},
\end{align}
where $f(x)$ is the Fermi-Dirac distribution function of the fermionic bath, and $\gamma_i$ are the Majorana fermions defined in the Hilbert space of the system.
The model parameters are the excitation energy of the Bogoliubov fermion $\epsilon\in\mathbb{R}$, and system-environment couplings $a,b\in \mathbb{C}$ whose relative ratio depends on the details of the Bogoliubov fermion. 
The rapidity spectrum is given by the eigenspectrum of the following non-Hermitian matrix $Z$ defined in Eq. (\ref{znonherm}):
\begin{align}
    Z=-H-i\mathrm{Re}M=-\frac{\epsilon}{4}\tau_y-ip\tau_x
    -iq\tau_z
    -ir\tau_0,\label{majoranarapid}
\end{align}
where $p=\mathrm{Re}[a^*b]$, $q=(|a|^2-|b|^2)/2$, $r=(|a|^2+|b|^2)/2$, and $\tau_\mu$ are the Pauli matrices ($\mu=0$: identity matrix).
Remarkably, the non-Hermitian matrix $Z$ does not depend on the distribution function $f$.
The non-Hermitian matrix (\ref{majoranarapid}) has the same form as the non-Hermitian Hamiltonian (\ref{majorananonham}) except for an additional Pauli matrix term.
The rapidity spectrum is given by
\begin{align}
    E_{\pm}=\pm\sqrt{(\epsilon/4)^2-(p^2+q^2)}-ir.\label{epm}
\end{align}
Since $\epsilon,p,q,r\in\mathbb{R}$, $E_+=-E^*_-$ holds, or both of them are pure imaginary numbers, which is a consequence of the particle hole symmetry. In the former case, where the dissipation is sufficiently small with respect to the excitation energy of the Bogoliubov fermion, the eigenvectors correspond to the ``complex fermions" in the $third$ $quantization$. On the other hand in the latter case, where the dissipation is large, they correspond to the ``Majorana fermions" with Re$E=0$ in the $third$ $quantization$. The transition point of these two phases is at the exceptional point $\epsilon/4=\sqrt{p^2+q^2}$, where the matrix $Z$ is not diagonalizable [Fig.\ref{fig4} (a)].

The transition in the above is regarded as a topological phase transition in the class D$^\dagger$ with the particle-hole unitary matrix $C=\tau_0$.
Because of PHS, ${\rm det}[i(Z-E_{\rm P})]$ takes only a real value if the reference energy $E_{\rm P}$ for the point gap is imaginary. 
Then, by choosing $E_{\rm P}$ as $-ir$, {\it i.e.} the energy at the exceptional point, we can captures the transition at the exceptional point as a topological phase transition.
When the determinant 
\begin{align}
{\rm det}[i(Z+ir)]=(\epsilon/4)^2-(p^2+q^2)    
\end{align}
is positive (negative), the system is topologically trivial (non-trivial). 
The nontrivial topology detects the anomalous zero modes with ${\rm Re}E=0$, {\it i.e.}, the Majorana zero modes.
In general, as rapidity spectra of general quadratic open Fermi systems obey the AZ$^\dagger$ symmetry, the non-Hermitian topology detects anomalous gapless modes of the $third$ $quantized$ fermions (Table \ref{table3}). However, it does not directly determine the properties of the steady state, which is implied by the fact that $Z$ of the above system does not depend on the distribution function $f$.
In the following, we investigate the entanglement structure of the above system.

\subsection{Qubit representation and entanglement }
To consider the entanglement, it is convenient to introduce the qubit representation.
The Hilbert space of the system is spanned by the basis vectors $|0\rangle$ and $|1\rangle:=\alpha^\dagger|0\rangle$, where $|0\rangle$ is the Fock vacuum of the Bogoliubov fermion $(\alpha,\alpha^\dagger)$. In this basis, the density matrix of the system and the system-bath coupling $c$ are expressed as
\begin{align}
    \rho=
    \begin{pmatrix}
    \rho_{00}&\rho_{01}\\
    \rho_{10}&\rho_{11}
    \end{pmatrix},~
    c=
    \begin{pmatrix}
    0&x\\
    y&0
    \end{pmatrix},
\end{align}
where $x=a+ib$ and $y=a-ib$.
Then the Liouvillian eigenvalue equation (\ref{eigenproblem}) of this system is given by
\begin{widetext}
\begin{align}
    \begin{pmatrix}
    -\left[2(1-f)|y|^2+2f|x|^2\right]&&&2(1-f)|x|^2+2f|y|^2\\
    &i\epsilon-(|x|^2+|y|^2)&2xy^*&\\
    &2yx^*&-i\epsilon-(|x|^2+|y|^2)&\\
    2(1-f)|y|^2+2f|x|^2&&&-\left[2(1-f)|x|^2+2f|y|^2\right]
    \end{pmatrix}
    \begin{pmatrix}
    \rho^{(i)}_{00}\\
    \rho^{(i)}_{01}\\
    \rho^{(i)}_{10}\\
    \rho^{(i)}_{11}
    \end{pmatrix}
    =\lambda_i
    \begin{pmatrix}
    \rho^{(i)}_{00}\\
    \rho^{(i)}_{01}\\
    \rho^{(i)}_{10}\\
    \rho^{(i)}_{11}
    \end{pmatrix}.\label{liuofmajorana}
\end{align}
\end{widetext}
The Liouvillian spectrum is calculated as $\{0,\pm\sqrt{-\epsilon^2+(4p)^2+(4q)^2}-4r,-8r\}$.
In the $third$ $quantization$ language, these eigenvalues correspond to the ``states" $\rho_{\rm ss},~\bar{\beta}_+^{\dagger}\rho_{\rm ss},~\bar{\beta}_-^{\dagger}\rho_{\rm ss}$, and $\bar{\beta}_+^{\dagger}\bar{\beta}_-^{\dagger}\rho_{\rm ss}$, and are consistent with Eqs. ($\ref{rapliurel}$) and (\ref{epm}). The Liouvillian is block-diagonalized into the ($\rho_{00},\rho_{11}$) and ($\rho_{01},\rho_{10}$) spaces, which comes from the fact that the quadratic open Fermi system preserves the parity of the $third$-$quantized$ fermions \cite{Prosen-2008}.

Next, we analyze the entanglement structure of the steady state.
The explicit form of the steady state ($\lambda=0$) can be obtained by solving the eigenvalue equation (\ref{liuofmajorana}):
\begin{align}
    \rho_{\rm ss}=
    \begin{pmatrix}
    F_1(x,y)&0\\
    0&F_2(x,y)
    \end{pmatrix},
\end{align}
where $F_1(x,y)=\left[(1-f)|x|^2+f|y|^2\right]/(|x|^2+|y|^2)$ and $F_2(x,y)=\left[(1-f)|y|^2+f|x|^2\right]/(|x|^2+|y|^2)$. We use the property of the density matrix Tr$[\rho]=1$.
The corresponding von Neumann entanglement entropy is
\begin{align}
S=-\sum^2_{i=1}F_i(x,y)\log F_i(x,y).
\end{align}
As expected, the entanglement of the steady state depends on the bath distribution function, while the rapidity spectrum does not.
To get rid of the thermal contribution, we set $f(\epsilon)=0$ or $1$.
The entanglement entropy $S$ takes the largest value $\log2$ at $|x|=|y|$.
This is the same value as the entanglement entropy of one qubit subsystem of the Bell states, the maximally entangled states of the two-qubit system.

\subsection{Topological Majorana fermion vs. steady-state Majorana fermion}
Under the strong dissipation limit $|x|,|y|\gg\epsilon$, the steady state is similar to a topological edge state.
To see this, we here review the Kitaev chain \cite{Kitaev-01}, a typical model of a topological superconductor in one dimension:
\begin{align}
    H_{\rm Kitaev}=\frac{-i}{2}\left(v\sum^{n}_{i=1}\gamma_{2i-1}\gamma_{2i}+w\sum^{n-1}_{i=1}\gamma_{2i}\gamma_{2i+1}      \right),
\end{align}
where $\gamma_{2i-1}=\alpha_i+\alpha^\dagger_i$ and $\gamma_{2i}=i(\alpha_i-\alpha^\dagger_i)$ with $(\alpha_i,\alpha^\dagger_i)$ being the complex fermion at the site $i$, $v>0$ is the onsite potential, $w>0$ is the hopping energy, and $n$ is the number of sites.
This model is topologically nontrivial for $v<w$, and there arise unpaired Majorana degrees of freedom $\gamma'_1,\gamma'_2$ at the different ends.
The low-energy effective theory of the nontrivial phase restricted in the lowest two states, which are protected by the large gap, is described by the unpaired Majorana fermions:
\begin{align}
    H_{\rm eff}=-i\frac{\epsilon}{2} \gamma'_1 \gamma'_2,
\end{align}
where $\epsilon$ is the small gap between the ground state and the first excited state.
In the nontrivial phase under the thermodynamic limit ($n\rightarrow\infty$), $\epsilon$ becomes 0, and the ground states are doubly degenerated with a finite excitation gap. For $v=0$ and $w\neq0$, these properties are unchanged even for $n=2$.
In the qubit representation, the four eigenstates of the $n=2$ Kitaev chain are given by the Bell states,
\begin{align}
    &\frac{1}{\sqrt{2}}(|0\rangle|0\rangle+|1\rangle|1\rangle),
    ~\frac{1}{\sqrt{2}}(|0\rangle|1\rangle+|1\rangle|0\rangle),\notag\\
    &\frac{1}{\sqrt{2}}(|0\rangle|0\rangle-|1\rangle|1\rangle),
    ~\frac{1}{\sqrt{2}}(|0\rangle|1\rangle-|1\rangle|0\rangle).
\end{align}
The former two states are the degenerated ground states, and the latter ones are the excited states with the gap $w$. Note that both of the ground states are maximally entangled states in the position space, $S=\log2$ at sites 1 and 2, and they are in the different sectors of the parity of the fermion number, which is the conserved quantity in the superconductor. 
In the topological quantum computation, the fermion parity is changed under the braiding process of the Majorana fermions at the ends of the topological superconducting nanowires \cite{Kitaev-01,Nayak}. 
The entanglement entropy at the boundary is a signal of the short-range entangled nature of the symmetry-protected topological phase.

Now we are in a position to compare the strong dissipation limit of the steady state with the topological edge states.
For simplicity, let us consider the case where $x=y\in\mathbb{R}$ and $f(\epsilon)=0$.
The entanglement entropy of the steady state is again $S=\log2$, which is the same value as that of the edge states of topological superconductors. For $2|a|^2>\epsilon$, $\lambda=\sqrt{-\epsilon^2+(2|a|^2)^2}-2|a|^2$ corresponds to the Liouvillian excitation that has the longest lifetime $1/\mathrm{Re}\lambda$.
In the large dissipation limit ($2|a|^2\gg\epsilon$), $\lambda$ becomes close to 0, which leads to the steady-state quasi degeneracy [Fig.\ref{fig4} (b)].
The corresponding ``eigenvector" with the longest lifetime of the Liouvillian superoperator is given by
\begin{align}
    \rho^{+}\sim
    \begin{pmatrix}
    0&1\\
    1&0
    \end{pmatrix}.
\end{align}
Remarkably, it mixes the different sectors of the fermion parity.
This fact implies that the presence of the quasi steady state is a resource of changing the fermion parity of the system under the perturbation such as Majorana braiding process discussed above, though the fermion parity of the total system is masked in the Lindblad formalism.
These results in the open quantum formalism would be helpful for deeper understanding of the Andreev vs. Majorana controversy.

\subsection{Example}
In general, the mismatch of the particle-hole ratio between the system and environment fermions is the source of the steady-state quantum entanglement. In the above model, this ratio is given by hand, though it is a function of experimentally controllable parameters in actual situations.
We here consider a simple system described by the following non-Hermitian matrix $Z$:
\begin{align}
    Z=-\frac{\epsilon}{4}\sigma_0\tau_y-\frac{\Delta}{4}\sigma_y\tau_x-\frac{h}{4}\sigma_x\tau_y-i\frac{g}{2}(\sigma_z\tau_0+\sigma_0\tau_0),
\end{align}
where $\epsilon$ is the electron energy, $\delta>0$ is the $s$-wave superconducting paring, $h>0$ is the magnetic field, and $\sigma_\mu$ and $\tau_\mu$ are Pauli matrices in the spin and particle-hole spaces, respectively.
We consider the spin-dependent dissipation $g>0$ such as a ferromagnetic metal and a half metal.
To consider the quantum entanglement, we set $f=0$.
For simplicity, we consider $\epsilon=0$.
The rapidity spectrum is given by 
\begin{align}
    E=\pm \frac{i}{2}\sqrt{g^2-\left(\frac{\Delta\pm h}{2}\right)^2}-\frac{ig}{2}.
\end{align}
The non-Hermitian topological transition occurs at $g=|\Delta\pm h|$.
Under the large dissipation limit, the ratio between the longest and the second longest lifetimes becomes $[(\Delta+h)/(\Delta-h)]^2$. 

The dissipation-dependence of the entanglement entropy $S$ is plotted in Fig.\ref{fig4} (c).
Under small but finite dissipation, the entanglement is about $2\log2$, which is twice as large as that of the Majorana edge mode discussed above.
Under large-dissipation limit, $S$ becomes close to $\log2$ for $h/\Delta=1.1$, while it is slightly different from $\log2$ for $h/\Delta=2.0$. In the former case, the second longest lifetime is too far from the longest one to affect the steady-state entanglement. In this sense, the former case is more close to the model treated in the previous subsections.

\section{Non-Hermitian topology and entanglement in one- and higher-dimensional open systems \label{onedim}}
In this section, we analyze the Lindblad equation whose rapidity spectrum has one-dimensional non-Hermitian topology. 
We again choose an appropriate setup and relate it to the chiral edge state of the Chern insulator in terms of the entanglement spectrum.
We also propose several realistic setups including higher-dimensional systems.

\subsection{Rapidity spectrum and chiral fermion}
\begin{table}[]
\caption{One-dimensional non-Hermitian topology and its physical interpretation.
The AZ$^\dagger$ classification is equivalent to the AZ Hermitian classification of the anomalous gapless fermion mode \cite{teo-kane}.
}
\label{table4}
\centering
$$
\begin{array}{ccl}
\mbox{~Symmetry~}&\mbox{~Classification~}&\mbox{~Interpretation~}\\
\hline \hline
{\rm A}&\mathbb{Z}&\mbox{~Chiral Dirac~}\\
{\rm D}^\dagger&\mathbb{Z}&\mbox{~Chiral Majorana~}\\
{\rm DIII}^\dagger&\mathbb{Z}_2&\mbox{~Helical Majorana~}\\
{\rm AII}^\dagger&\mathbb{Z}_2&\mbox{~Helical Dirac~}\\
{\rm C}^\dagger&2\mathbb{Z}&\mbox{~Chiral Dirac~}\\
\hline
\hline
\end{array}
$$
\end{table}

\begin{figure*}[]
\begin{center}
　　　\includegraphics[width=16cm,angle=0,clip]{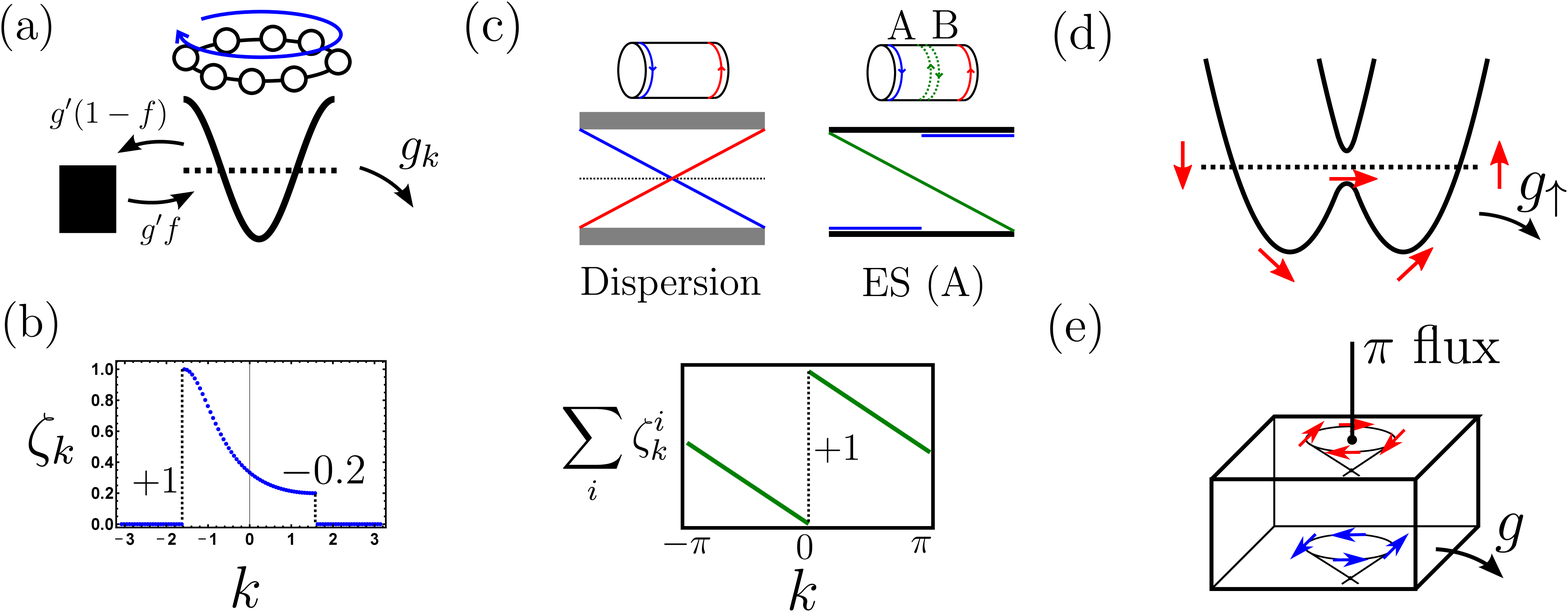}
　　　\caption{(a) Schematic picture of one-dimensional open Fermi system with class-A non-Hermitian topology. One environment is a thermal bath with distribution function $f(\epsilon)$, and the other environment is the momentum-dependent dissipator whose chemical potential is much lower than that of the system. The rapidity spectrum is described by the Hatano-Nelson model. (b) Entanglement spectrum (ES) of $n=100$ one-dimensional open Fermi system (\ref{entanglementspe}). (c) Dispersion, ES, and trace of ES of Chern insulator with $C=+1$. The discontinuity of the ES detects the nontrivial Chern number. (d) Schematic picture of the spin-momentum-locked band and spin-dependent dissipation. Even the coupling constant $g_{\uparrow}$ is $k$-independent, dissipation is effectively $k$-dependent owing to the spin-momentum locking. (e) Schematic picture of the two-dimensional class-AII$^\dagger$ skin effect in surface states of a topological insulator.}
　　　\label{fig5}
\end{center}
\end{figure*}

As the simplest example, we consider an $n$-site one-dimensional open Fermi system whose rapidity spectrum is described by the Hatano-Nelson model (\ref{hatano}) under the PBC.
Suppose that the system is translation invariant.
Then the $2^n\times2^n$ density matrix can be decomposed into $2\times 2$ sectors labeled by momentum $k$:
\begin{align}
    \rho=\prod_{k}\rho_k.
\end{align}
We also assume that the system is coupled with two environments: a momentum-dependent dissipator and a thermal bath [Fig.\ref{fig5} (a)], and the dynamics of the density matrix is completely decomposed into momentum sectors: 
\begin{align}
    \frac{d\rho_k}{dt}=&-i[H_k,\rho_k]+\left[2c_k\rho_k c^\dagger_k-\{c^\dagger_kc_k,\rho_k\}\right]\notag\\
    &+(1-f(\epsilon_k))\left[2d\rho_k d^\dagger-\{d^\dagger d,\rho_k\}\right]\notag\\
    &+f(\epsilon_k)\left[2d^\dagger\rho_k d-\{dd^\dagger,\rho_k\}\right],\\
    c_k=&\frac{\sqrt{g}}{2}(1+ie^{-ik})(\gamma_{k,1}-i\gamma_{k,2}),\\
    d=&\sqrt{g'}(\gamma_{k,1}-i\gamma_{k,2}),\\
    \gamma_{k,1}=&\alpha_k+\alpha_k^\dagger,~\gamma_{k,2}=i(\alpha_k-\alpha_k^\dagger),\\
    H_k=&-i\frac{\epsilon_k}{2} \gamma_{k,1}\gamma_{k,2}=\epsilon_k \alpha^\dagger_k\alpha_k+\mathrm{const.},
\end{align}
where $\epsilon_k=2t\cos k$ with $t\in\mathbb{R}$ being the hopping energy, $f(x)$ is the Fermi distribution function of the thermal bath, and $g,~g'>0$ are the system-dissipator and system-bath couplings, respectively.
For simplicity, we have assumed the chemical potential of the dissipator is much lower than that of the system, and the momentum-dependent dissipation does not contain the distribution function of the dissipator.
The rapidity spectrum is given by the eigenspectrum of the non-Hermitian matrix $Z_k$:
\begin{align}
    Z_k=-\frac{2t\cos k}{4}\tau_y-i\left[\frac{g}{2}(1+\sin k)+g'\right]\tau_0.
\end{align}
Thus, the rapidity spectrum is given by the stack of two Hatano-Nelson models with opposite winding numbers:
\begin{align}
    E_{k,\pm}=\frac{\pm 2t\cos k-i~2g\sin k}{4}-i\left(g'+\frac{g}{2}\right).
\end{align}
The system has the U(1) symmetry and is essentially in the class A, and the class-A non-Hermitian topology can be separately defined for particle and hole sectors. 

In general, one-dimensional non-Hermitian topology detects the anomalous gapless fermion modes (Table \ref{table4}). In the present case, the class-A spectral curve indicates the presence of the $third$ $quantized$ chiral fermion. The class-A non-Hermitian topology also predicts the skin effect as noted in Sec. \ref{physint}.
In the following, we analyze the entanglement structure of this system and its physical consequence.

\subsection{Qubit representation and entanglement spectrum}

To consider the entanglement, we again introduce the qubit representation for each momentum.
The Hilbert space of the momentum-resolved system is spanned by the basis vectors $|0\rangle$ and $|1\rangle:=\alpha^\dagger_k|0\rangle$, where $|0\rangle$ is the Fock vacuum of the fermion $(\alpha,\alpha^\dagger)$.
Then the eigenvalue equation (\ref{eigenproblem}) for each momentum is given by
\begin{widetext}
\begin{align}
    \begin{pmatrix}
    -8g'f(\epsilon_k)&&&8g'[1-f(\epsilon_k)]+2g_k\\
    &i\epsilon_k-(4g'+g_k)&&\\
    &&-i\epsilon_k-(4g'+g_k)&\\
    8g'f(\epsilon_k)&&&-\left[8g'\{1-f(\epsilon_k)\}+2g_k\right]
    \end{pmatrix}
    \begin{pmatrix}
    \rho^{(i)}_{k,00}\\
    \rho^{(i)}_{k,01}\\
    \rho^{(i)}_{k,10}\\
    \rho^{(i)}_{k,11}
    \end{pmatrix}
    =\lambda_{k,i}
    \begin{pmatrix}
    \rho^{(i)}_{k,00}\\
    \rho^{(i)}_{k,01}\\
    \rho^{(i)}_{k,10}\\
    \rho^{(i)}_{k,11}
    \end{pmatrix},
\end{align}
\end{widetext}
where $g_k:=g(2+2\sin k)$.
The Liouvillian spectrum is calculated as $\{0,\pm2t\cos k-i(4g'+g_k),-2i(4g'+g_k)\}$.
The explicit form of steady-state density matrix ($\lambda_k=0$) is given by
\begin{align}
    \rho_{k,\mathrm{ss}}=\frac{1}{4g'+g_k}
    \begin{pmatrix}
    4g'[1-f(\epsilon_k)]+g_k&0\\
    0&4g'f(\epsilon_k)
    \end{pmatrix}.
\end{align}
To investigate the quantum entanglement, we consider the zero temperature and set $f(x)=\theta(-x)$, where $\theta$ is the step function. In the quadratic Fermi system, one convenient tool to express the quantum entanglement is the single-particle entanglement spectrum, which is the occupation number in a single-particle level.
In this case, the momentum-resolved entanglement eigenvalue is given by
\begin{align}
    \zeta_k=\rho^{11}_{k,\mathrm{ss}}=\frac{4g'\theta(-\epsilon_k)}{4g'+g_k}.\label{entanglementspe}
\end{align}
The entanglement spectrum for $t<0$ and $g/g'=4$ is plotted in Fig.\ref{fig5} (b).

\subsection{Chern insulator vs. steady state}
Under the strong dissipation limit $g\gg g'$, the entanglement structure of the steady state is similar to that of the Chern insulator, which is a typical topological insulator in the class A.
Reference \cite{Alexandradinata} has characterized the Chern insulator in terms of the entanglement spectrum.
The Chern insulators are classified by the first Chern number $C_1\in \mathbb{Z}$, which is the two-dimensional class-A topological number. Under the OBC in $x$ direction and the PBC in $y$ direction, 
the Chern number counts the number of chiral edge modes with negative (positive) chirality at the left (right) boundary.
Let us consider the $C_1=1$ Chern insulator with the same boundary condition
and divide it into two parts (A and B).
A typical dispersion of the total system 
and the corresponding entanglement spectrum defined in the region A
are shown in Fig.\ref{fig5} (c).
The occupied number of the bulk states is constant with respect to the momentum, while that of the edge state at the left boundary is discontinuous at the Fermi momentum.
Remarkably, there arises a mode at the boundary between A and B that connects the bulk occupied states and unoccupied states.
This mode indicates the short-range entangled nature of the (U(1)) symmetry-protected topological phase and cannot be gapped out without closing the bulk gap.
Reference \cite{Alexandradinata} has proposed the trace index that counts the 
\begin{align}
    \mathcal{A}_{U(1)}:=&\lim_{n_x,n_y\rightarrow\infty}\sum_{K_{\rm cross},i}\zeta^i_{K_{\rm cross}+\pi/N_y}-\zeta^i_{K_{\rm cross}-\pi/N_y}\notag\\
    =&\sum_{K_{\rm cross}}n_{(-)}(K_{\rm cross})-n_{(+)}(K_{\rm cross}),
\end{align}
where $K_{\rm cross}$ are the Fermi momenta, $n_{(\pm)}$ is the number of left-edge states at $K_{\rm cross}$ with positive/negative chirality.
Reference \cite{Alexandradinata} has shown the equivalence between the trace index and the Chern number:
\begin{align}
    \mathcal{A}_{U(1)}=C_1.
\end{align}
This formula holds for the present case [Fig.\ref{fig5} (c)].
In summary, the trace index defined in the subsystem detects the symmetry-protected topology of the Chern insulator.

In the following, we introduce the trace index to the one-dimensional open Fermi system. The trace index for Eq. (\ref{entanglementspe}) is given by
\begin{align}
    \mathcal{A}_{U(1)}&=\lim_{\delta\rightarrow0}\sum_{K_{\rm cross}=\pm \pi/2}\xi_{K_{\rm cross}+\delta}-\xi_{K_{\rm cross}-\delta}\notag\\
    &=1-\frac{g'}{g'+g}.
\end{align}
Unlike an equilibrium insulator, the trace index is not quantized in general open Fermi systems.
However, it becomes close to unity under the large dissipation limit $g\gg g'$, while it becomes 0 for trivial non-Hermitian topology ($g=0$). In this sense, the entanglement spectrum of the steady state with a class-A nontrivial rapidity spectrum behaves as if the mixed state of a half subsystem of the Chern insulator. Since the trace index has been defined in other classes such as in the $\mathbb{Z}_2$ time-reversal topological insulator \cite{Alexandradinata}, similar relations should hold for other AZ$^\dagger$ symmetries.

\subsection{Remark: nonreciprocal current generation and skin effect}
In the above model, the asymmetry of $\zeta_k$ with respect to $k$ and $\epsilon_k$ causes the nonreciprocal fermionic $U(1)$ and thermal currents:
\begin{align}
    J=&\int^{\pi/2}_{\pi/2}\frac{dk}{2\pi}\zeta_k\frac{d\epsilon_k}{dk},\\
    J_E=&\int^{\pi/2}_{\pi/2}\frac{dk}{2\pi}\zeta_k\epsilon_k\frac{d\epsilon_k}{dk}.
\end{align}
The non-equilibrium nature comes from the implicit ``voltage" between the thermal bath and the momentum-dependent dissipator whose chemical potential is lower than the band bottom of the one-dimensional system. 
In this sense, the current circularly flows in the transverse direction to the applied voltage. This example indicates that entanglement structures can be used as the source of the nonreciprocal current.

If we impose the OBC to the same model, the above current cannot exist in the presence of the boundaries, which leads to the fermionic accumulation at the boundaries and generates the voltage between the right and left boundaries in the steady state. This phenomenon might be related to the fact that the rapidity spectrum suffers from the class-A skin effect, though the detailed analysis under the OBC is beyond the scope of this paper.

In other classes and dimensions whose non-Hermitian topology detect the symmetry-protected or higher-dimensional skin effects discussed in the section \ref{physint}, 
other types of nonequilibrium fermionic current can be discussed.
In the one-dimensional class AII$^\dagger$, the time-reversal symmetric current such as the spin current flows under the PBC, and the corresponding accumulation such as the spin accumulation occurs under the OBC.
In higher dimensions, the current flows in the radial direction in the presence of a topological defect, which causes the particle accumulation under the OBC. 

Note that in the absence of the applied voltage, the current flow under the PBC and the fermionic accumulation under the OBC do not occur, which is the typical case where the nontriviality of the rapidity spectrum is not related to a nontrivial physical phenomenon in the steady state.
If we only consider the momentum-dependent dissipator and its chemical potential is set to be 0 in the considered system, the model is equivalent to the Hermitian correlated system with momentum-dependent self-energy. Such a case has been investigated in Ref. \cite{Okuma-correlated}.

\subsection{Example: on-site dissipation in multi-band system}
The Hatano-Nelson model requires the asymmetric hopping term in real space, or equivalently, the momentun-dependent dissipation in momentum space. An implementation of such a term is a nontrivial task in realistic condensed matter physics.
However, the momentum-dependent dissipation can be effectively realized even under the onsite dissipation in multi-band systems \cite{YSW-18-Chern,KSU-18,Yi-Yang-20,Bessho-Sato-20}.
We here give an intuitive interpretation of this phenomenon.

The trick is similar to a topological superconductor in which the $p$-wave paring is effectively realized by the combination of the $s$-wave paring and spin-momentum-locked band \cite{Sato-Ando,Sato-Fujimoto,Lutchyn,O-R-vO}. 
As an example, we here consider the two-band Hamiltonian with spin degree of freedom $\sigma$:
\begin{align}
    H_k=\epsilon_k\sigma_0+\bm{d}_k\cdot\bm{\sigma},
\end{align}
where $\bm{d}_{k}=(d^x_k,d^y_k,d^z_k)$ is the spin-dependent interaction. The band dispersion is given by $E_{k,\pm}=\epsilon_k\pm |\bm{d}_k|$. The spin polarization of each mode ($k,\pm$) is given by $\pm\bm{d}_k/(2|\bm{d}_k|)$. If the spin-rotational symmetry is completely broken, the spin polarization can depend on the momentum. For example, $\bm{d}_k=(h,0,\lambda k)$, the combination of a Rashba spin-orbit interaction and a time-reversal symmetry-breaking term, which is the same system as the topological superconducting quantum nanowire \cite{Lutchyn,O-R-vO} except for superconducting proximity, induces the spin-texture in both bands that varies continuously in momentum [Fig.\ref{fig5} (d)]. For such spin-momentum-locked bands, the spin-dependent momentum-independent dissipation acts as the effectively momentum-dependent dissipation.
Under such a dissipation, the rapidity spectrum can have a nontrivial winding number, which induces the nonreciprocal current under the PBC and fermion accumulation under the OBC.

\subsection{Example: chirality-dependent dissipation in Dirac systems}
In Refs. \cite{Lee-Vishwanath-19,Bessho-Sato-20,OKSS-20}, several massless Dirac systems that have AZ$^\dagger$ non-Hermitian topology, including the model $(\ref{2dskin})$, have been investigated. As in the case of the model $(\ref{2dskin})$, the valley-dependent dissipation can induce the non-Hermitian topology.
The surface states of the topological insulators and Graphene are good candidates of two-dimensional AII$^\dagger$ non-Hermitian topological systems.
In the former case, the Dirac cones with opposite chiralities appear on the top and bottom surfaces, which enables one to connect only one of them to the dissipator [Fig.\ref{fig5} (e)].
In the latter case, the Dirac cones with different chiralities cannot be separated in real space.
Thus for the valley-dependent dissipation, valleytronic techniques are required.
Under the insertion of the $\pi$ flux, there occurs the anomalous fermion production and current generation in radial direction, which causes two-dimensional $\mathbb{Z}_2$ skin effect discussed in Sec. \ref{physint}.

\section{Discussion \label{discussion}}
In this section, we discuss several related topics.
\subsection{Transient Liouvillian dynamics}
While the main target of this paper is the steady state, the rapidity and Liouvillian spectra are often discussed in terms of the dynamics, and they are believed to convenient notion for the description of the dynamics.
In our previous paper \cite{Okuma-Sato-20}, however, we have pointed out that the dynamical properties are also affected by the pseudospectrum.
We here remark this point and relate it to recent papers.

Mathematically, $\epsilon$ pseudospectrum is defined as
\begin{align}
    \sigma_{\epsilon}(H)=\{E \in\mathbb{C}|\ \|(H-E)|v\rangle \|<\epsilon\ {\rm for\ some}\ |v\rangle\},
\end{align}
where $H$ is a matrix, and $|v\rangle$ is a normalized vector.
By definition, the pseudospectrum is almost an eigenvalue, and is not so important in conventional Hermitian physics.
In fact, in the case of normal matrices (i.e., $[H,H^\dagger]=0$) such as Hermitian matrices, the $\epsilon$ pseudospectrum is just the $\epsilon$ neighborhood of the spectrum. However, it is larger than the $\epsilon$ neighborhood in the case of nonnormal matrices.
The pseudospectrum becomes an important factor in the transient dynamics of various linear equations, while the long-time dynamics is completely captured in terms of the exact spectrum \cite{Trefethen}.
Thus the Liouvillian dynamics should be affected by the pseudospectrum of the Liouvillian superoperator in the transient region.

The larger the nonnormality of the matrix is, the more important the pseudospectrum is.
In particular, the following relationship holds for a translation-invariant class-A matrix $H$ under the OBC \cite{Trefethen, Okuma-Sato-20,OKSS-20}:
\begin{align}
    \lim_{\epsilon\rightarrow0}\lim_{n\rightarrow\infty}\sigma_\epsilon(H)=\sigma_{\rm SIBC}(H),\label{sibc}
\end{align}
where $\sigma_{\rm SIBC}(H)$ is the spectrum under the semi-infinite boundary condition, in which there is only one open boundary. According to the index theorem \cite{Bottcher,Trefethen,OKSS-20,Okuma-Sato-20}, $\sigma_{\rm SIBC}(H)$ is given by the PBC spectrum together with the region enclosed by the PBC curve with nonzero winding. Thus, under typical class-A non-Hermitian topology, the region of the pseudospectrum is much larger than the $\epsilon$ neighborhood of the spectrum. Similar statements hold for other classes such as the class AII$^\dagger$ \cite{Okuma-Sato-20}.
Exceptional points are also the resource of the nonnormal pseudospectral behavior.

Reference \cite{haga-20} has investigated the dynamics of the Liouvillian skin effect \cite{Song-yao-wang-19}.
They have claimed that the lifetime is not captured by the Liouvillian spectrum. This might be because their definition of the lifetime captures the transient region and does not the long-time dynamics. In this sense, this phenomenon might be closely related to the notion of the pseudospectrum.
Reference \cite{mori-20} has also investigated the similar physics in different situations, and discussed the time-scale-dependent definition of the lifetime, which might be useful to distinguish the pseudospectral and spectral behaviors.

\subsection{Non-Hermitian model as lattice implementation of quantum anomaly}
It is well known that quantum field theories with quantum anomaly cannot be implemented in the pure lattice systems.
However, such theories can be implemented by allowing the non-Hermiticity in some sense.
In terms of spectral theories, Refs. \cite{Lee-Vishwanath-19,Bessho-Sato-20} have shown that non-Hermitian topology is related to the quantum anomaly.
Moreover, the present paper has pointed out that the non-Hermitian skin effects are consequences of the anomaly-induced transport including the anomalous fermion production induced by the topological defects in higher dimensions.

In terms of quantum entanglement, the present paper has investigated that dissipation-induced steady state in open Fermi systems can have similar entanglement entropy and spectrum as that of the edge states of topological insulators and superconductors.
These statements imply that the quantum field theories with quantum anomaly can be investigated by introducing the non-Hermiticity or the open quantum nature.

Since quantum anomaly is also discussed in interacting systems, non-Hermitian topological phenomena would be generalized to interacting systems by developing the above ideas. In particular, the anomaly-induced transport in interacting systems indicates the presence of the skin effect in the interacting non-Hermitian systems.
Such models could be constructed from the dissipation that induces the imbalance of the lifetime of the quantum anomalous modes with different topological number.

\acknowledgements
This work was supported by JST CREST Grant No.~JPMJCR19T2, Japan. N.O. was supported by KAKENHI Grant No.~JP20K14373 from the JSPS. M.S. was supported by KAKENHI Grant No.~JP20H00131 from the JSPS.

\appendix
\section{Topological aspect of class-A non-Hermitian skin effect \label{appendixa}}
In this section, we briefly review the relationship between the winding number and the class-A non-Hermitian skin effect \cite{OKSS-20}.
The conventional (class-A) non-Hermitian skin effect, where the OBC spectrum is drastically different from the PBC one, can occur for one-dimensional non-Hermitian tight-binding models without any specific symmetry (essentially class A).
If a one-dimensional class-A non-Hermitian Hamiltonian with finite-range hoppings $H$ has the translation invariance in the bulk, the following theorem holds in the infinite-volume limit:\\

{\bf Theorem I}~~The OBC spectrum $\sigma(H_{\rm OBC})$ cannot have nonzero winding number. Consequently, if the PBC spectral curve $\sigma(H_{\rm PBC})$ has a nonzero winding number, the non-Hermitian skin effect inevitably occurs.\\
\\
Here the winding number of the spectral curve $C$ around $E\in\mathbb{C}$ $W(E)$ is defined as
\begin{equation}
W \left( E \right)
:= \oint_{C} \frac{d\beta}{2\pi i} \frac{d}{d\beta} \log \left( H \left( \beta \right) - E \right),\label{eq: winding number}
\end{equation} 
where $H(\beta)$ is given by replacing the Bloch wave factor $e^{ik}$ of the Bloch Hamiltonian $H(e^{ik})$ with the non-Bloch wave factor $\beta\in\mathbb{C}$.

In the following, we give a brief sketch of the proof of the Theorem I. 
The main difficulty to compare the OBC and PBC spectra comes from the absence of the simple analytic properties of the OBC spectrum.
To avoid the direct comparison, we introduce the semi-infinite boundary condition, where the boundary is only at the left-hand side of the system, and compare it with the PBC and OBC. The semi-infinite spectrum is characterized in terms of the PBC curve by the index theorem in spectral theory: \cite{Trefethen,Bottcher}:
\\

{\bf Theorem II}~~
The spectrum of semi-infinite system $\sigma(H_{\rm SIBC})$ is equal to the PBC spectral curve $\sigma \left( H_{\rm PBC} \right)$ together with the whole area of $E \in \mathbb{C}$ enclosed by $\sigma \left( H_{\rm PBC} \right)$ with $W \left( E\right) \neq 0$. For $W \left( E \right) < 0$ ($W \left( E \right) > 0$), $\ket{E}$ is a right (left) eigenstate of $H_{\rm SIBC}$ localized at the boundary [i.e., $H_{\rm SIBC} \ket{E} = E \ket{E}$ ($\bra{E} H_{\rm SIBC} = \bra{E} E$)].\\
\\
Remarkably, the semi-infinite spectrum is given by a two-dimensional region in the complex plane if the PBC curve has a nonzero winding number.
Roughly speaking, the OBC system has an additional boundary condition, the  
OBC spectrum is in the semi-infinite spectrum:
\begin{align}
    \lim_{n \to \infty}\sigma \left( H_{\rm OBC} \right)\subset\sigma(H_{\rm SIBC}).\label{eq: inclusion OBC-SIBC}
\end{align}
Here we explicitly write the infinite-volume limit $ \lim_{n \to \infty}$ with $n$ being the number of sites.
For a mathematical proof, see Refs. \cite{OKSS-20,Trefethen}.

A crucial step is the imaginary gauge transformation: $H_{\rm OBC} \rightarrow V_{r}^{-1} H_{\rm OBC} V_{r}$ and $H_{\rm SIBC} \rightarrow V_r^{-1} H_{\rm SIBC} V_r$, where $[V_r]_{i,j}=\delta_{ij}r^i$ with $r \in \left( 0, \infty \right)$. For each transformation, we still have the inclusion in Eq.~(\ref{eq: inclusion OBC-SIBC}):
\begin{equation}
\lim_{n \to \infty} \sigma \left( V_{r}^{-1} H_{\rm OBC} V_{r} \right) \subset \sigma \left( V^{-1}_{r} H_{\rm SIBC} V_{r} \right).
	\label{eq: inclusion OBC-SIBC-2}
\end{equation}
This imaginary gauge transformation does not change the spectrum of $H_{\rm OBC}$ because the imaginary gauge transformation is a similarity transformation, which preserves the spectrum of finite-dimensional matrices. However, it changes the spectrum of $H_{\rm SIBC}$. In fact, $H \left( k \right)$ changes to $H \left( k-\ii\log r \right)$ through $V_{r}$. Nevertheless, Eq.~(\ref{eq: inclusion OBC-SIBC-2}) implies that the transformed semi-infinite spectrum includes the spectrum of $H_{\rm OBC}$ for any transformation $V_{r}$. Thus, we have
\begin{equation}
\lim_{n \to \infty} \sigma \left( H_{\rm OBC} \right) \subset \bigcap_{r \in \left( 0, \infty \right)} \sigma \left( V^{-1}_{r} H_{\rm SIBC} V_{r} \right).
	\label{eq: inclusion OBC-SIBC-3}
\end{equation}
Because of Theorem~II, when the PBC curve has a nonzero winding number, right (left) boundary modes with eigenenergy $E$ appear in the semi-infinite system. Let us choose an appropriate imaginary gauge $V_{r}$ such that these boundary modes are transformed to delocalized bulk modes. Then, $E$ is on the edges of $\sigma \left( V_{r}^{-1} H_{\rm SIBC} V_{r}\right)$, whereas it is originally located inside $\sigma \left( H_{\rm SIBC} \right)$. Thus, the intersection of $\sigma \left( H_{\rm SIBC} \right)$ and $\sigma \left( V_{r}^{-1} H_{\rm SIBC} V_{r}\right)$ is strictly smaller than $\sigma \left( H_{\rm SIBC} \right)$. Repeating this procedure for all $V_{r}$ with $r \in \left( 0, \infty \right)$, the right-hand side of Eq.~(\ref{eq: inclusion OBC-SIBC-3}) reaches a curve with no winding, otherwise a contradiction arises. Since this region includes the OBC curve because of Eq.~(\ref{eq: inclusion OBC-SIBC-3}), the OBC spectrum is also a curve with no winding and different from the PBC curve with a nonzero winding number, which implies the inevitable occurrence of the non-Hermitian skin effect if the PBC curve has a nonzero winding.


%

\end{document}